\documentclass[12pt]{article}
\usepackage{amsmath}
\usepackage{pstricks,pst-plot,pst-grad,pstcol,pst-char}

\makeatletter

\@addtoreset{equation}{section}

\def\draftdate{\relax}
\def\mda{\relax}
\def\mua{\relax}
\def\mla{\relax}
\def\draft{
\def\thtystars{******************************}
\def\sixtystars{\thtystars\thtystars}
\typeout{}
\typeout{\sixtystars**}
\typeout{* Draft mode!
         For final version remove \protect\draft\space in source file *}
\typeout{\sixtystars**}
\typeout{}
\def\draftdate{\today}
\def\mua{\marginpar[\boldmath\hfil$\uparrow$]%
                   {\boldmath$\uparrow$\hfil}%
                    \typeout{marginpar: $\uparrow$}\ignorespaces}
\def\mda{\marginpar[\boldmath\hfil$\downarrow$]%
                   {\boldmath$\downarrow$\hfil}%
                    \typeout{marginpar: $\downarrow$}\ignorespaces}
\def\mla{\marginpar[\boldmath\hfil$\rightarrow$]%
                   {\boldmath$\leftarrow $\hfil}%
                    \typeout{marginpar: $\leftrightarrow$}\ignorespaces}
\overfullrule 5pt
\oddsidemargin -15mm
\marginparwidth 29mm
}

\def\stars{\strut\leaders\hbox{*}\hfill\strut}
\def\starline{\hfil\strut\hfil\hbox to \textwidth {\stars}\hfil}

\newcommand\Ref[1]     {Ref.\,\cite{#1}}
\newcommand\Refs[1]    {Refs.\,\cite{#1}}
\newcommand\eqn[1]     {Eq.\,(\ref{#1})}
\newcommand\eqns[2]    {Eqs.\,(\ref{#1}) and~(\ref{#2})}
\newcommand\eqnss[2]   {Eqs.\,(\ref{#1})--(\ref{#2})}
\newcommand\fig[1]     {Fig.\,{\ref{#1}}}
\newcommand\sect[1]    {Sect.\,{\ref{#1}}}

\newcommand\app[1]     {Appendix~\ref{#1}}
\newcommand\tab[1]     {Table~\ref{#1}}
\newcommand\nn         {\nonumber}
\newcommand\res[3]     {$(#1\pm #2)\cdot 10^{#3}$}

\def\beq{\begin{equation}}
\def\eeq{\end{equation}}
\def\beeq{\begin{eqnarray}}
\def\eeeq{\end{eqnarray}}
\newcommand\bom[1]     {{\mbox{\boldmath $#1$}}}
\def\aand{\!\!\!\!\!\!\!\!&&}

\newcommand\as         {\ensuremath{\alpha_{\mathrm{s}}}}
\newcommand\asb        {\ensuremath{\alpha_{\mathrm{s}}^{\mathrm u}}}
\newcommand\gs         {\ensuremath{g_{\mathrm{s}}}}
\newcommand\aeps{\frac{\as}{2\pi}\,S_\eps\left(\frac{\mu^2}{Q^2}\right)^{\eps}}

\newcommand{\CF}       {C_{\mathrm{F}}}
\newcommand{\CA}       {C_{\mathrm{A}}}
\newcommand{\TR}       {T_{\mathrm{R}}}
\newcommand{\Nc}       {N_{\mathrm{c}}}
\newcommand{\Nf}       {n_{\mathrm{f}}}
\newcommand{\Ns}       {n_{\mathrm{s}}}
\newcommand{\bT}       {\bom{T}}

\newcommand\qb         {{\bar q}}
\newcommand\msbar      {\ensuremath{{\overline {\rm MS}}}}
\newcommand\muR[1]     {\ensuremath{\mu_R^{#1}}}
\newcommand\e          {{\mathrm e}}

\renewcommand\O        {{\mathrm O}}
\newcommand\Oe[1]      {\ensuremath{\mathrm O(\eps^{#1})}}
\newcommand{\eps}      {\varepsilon}

\newcommand\Li         {\mathop{\mathrm{Li}}\nolimits}
\newcommand\Real       {\mathop{\mathrm{Re}}\nolimits}

\newcommand\ldot       {\!\cdot\!}
\newcommand\smfrac[2]  {{\textstyle\frac{#1}{#2}}}
\newcommand\hf         {\smfrac{1}{2}}

\newcommand{\rd}       {{\mathrm{d}}}
\newcommand{\PS}[1]    {\rd\phi_{#1}}
\newcommand\tsig[1]    {\sigma^{\mathrm{#1}}}
\newcommand\dsig[1]    {\rd\sigma^{{\rm #1}}}
\newcommand\dsiga[2]   {\rd\sigma^{{\rm #1,A}_{\scriptscriptstyle #2}}}
\newcommand{\Jac}[2]   {{\cal J}^{(#1)}_{#2;m}}

\newcommand\la         {\langle}
\newcommand\ra         {\rangle}

\newcommand{\cM}       {{\cal M}}
\newcommand{\cA}       {{\cal A}}
\newcommand\SME[3]     {|{\cal M}_{#1}^{(#2)}{(#3)}|^2}
\newcommand\M[2]       {\ensuremath{|{\cal{M}}_{#1}^{#2}|^2}}
\newcommand\bra[3]     {\la {\cal M}_{#1}^{#2}#3|}
\newcommand\ket[3]     {|{\cal M}_{#1}^{#2}#3\ra}

\newcommand{\mom}[1]   {\{p\}^{#1}}
\newcommand{\momt}[1]   {\{\ti{p}\}^{#1}}

\newcommand{\bA}[1]    {\bom{\mathrm A}_{#1}}
\newcommand{\bC}[1]    {\bom{\mathrm C}_{#1}}

\newcommand{\bS}[1]    {\bom{\mathrm S}_{#1}}

\newcommand{\bCS}[2]   {\bom{\mathrm C}_{#1}\bom{\mathrm S}_{#2}}

\newcommand{\bSCS}[1]  {\bom{\mathrm C}\kern-2pt\bom{\mathrm S}_{#1}}

\def\hP{\hat{P}}
\newcommand{\pole}    {{\cal P}\!oles\,}
\newcommand{\finite}   {{\cal F}\!in\,}
\newcommand{\calS}     {{\cal S}}
\newcommand{\bcA}[2]   {{\bom{\cal A}}_{#1}}%^{#2}}
\newcommand{\cC}[2]    {{\cal C}_{#1}^{#2}}
\newcommand{\cS}[2]    {{\cal S}_{#1}^{#2}}
\newcommand{\cCS}[3]   {{\cal C}_{#1}^{~}{\cal S}_{#2}^{#3}}

\newcommand{\cSCS}[1]  {{\cal C}\kern-2pt{\cal S}_{#1}^{~}}

\newcommand{\IcC}[2]   {{\mathrm C}_{#1}}%^{#2}}
\newcommand{\IcS}[2]   {{\mathrm S}_{#1}}%^{#2}}
\newcommand{\IcCS}[1]  {\mathrm{CS}}%^{#1}}
\newcommand{\R}[2]     {{\mathrm R}_{#1}}%^{#2}}
\newcommand{\bI}       {\bom{I}}
\newcommand{\bIt}      {{\bom{I}}}%{{\bom{\ti{I}}}}

\newcommand{\ti}[1]    {\tilde{#1}}
\newcommand{\wti}[1]   {\widetilde{\,#1\,}}

\newcommand\tzz[2]     {z_{#1,#2}}%{\tilde z_{#1,#2}}
\newcommand\kTm[1]     {k_{\perp}^{#1}}
\newcommand\kT[1]      {k_{\perp,#1}}
\newcommand\kTt[1]     {k_{\perp,#1}}%{\tilde{k}_{\perp,#1}}
\renewcommand\S        {{\scriptscriptstyle\rm S\!.}}
\newcommand\NS         {{\scriptscriptstyle\rm N\!.S\!.}}

\begin{document}

% Title page

\begin{titlepage}
\renewcommand{\thefootnote}{\fnsymbol{footnote}}
\begin{flushright}
hep-ph/0609043 
     \end{flushright}
\par \vspace{5mm}
\begin{center}
{\Large \bf A subtraction scheme for computing QCD jet cross sections
at NNLO: \\[.5em] 
regularization of real-virtual emission}
\end{center}

\par \vspace{2mm}
\begin{center}
{\bf G\'abor Somogyi}, {\bf Zolt\'an Tr\'ocs\'anyi},\\[.5em]
{University of Debrecen and \\Institute of Nuclear Research of
the Hungarian Academy of Sciences\\ H-4001 Debrecen, PO Box 51, Hungary}
\end{center}

\par \vspace{2mm}
\begin{center} {\large \bf Abstract} \end{center}
\begin{quote}
\pretolerance 10000
We present a subtraction scheme for computing jet cross sections
in electron-positron annihilation at next-to-next-to-leading order
accuracy in perturbative QCD. In this second part we deal with the
regularization of the real-virtual contribution to the NNLO correction.
\end{quote}

\vspace*{\fill}
\begin{flushleft}
September 2006
\end{flushleft}
\end{titlepage}
\clearpage

% Have a TOC even for drafts
\tableofcontents

\renewcommand{\thefootnote}{\fnsymbol{footnote}}

% Begin actual paper

%
% Introduction
%

\section{Introduction}
\label{sec:intro}

In recent years a lot of effort has been devoted to the extension of
the subtraction method of computing QCD corrections at the
next-to-leading order (NLO) accuracy to the computation of the radiative
corrections at the next-to-next-to-leading order (NNLO) 
\cite{Gehrmann-DeRidder:2004tv,Gehrmann-DeRidder:2004xe,Weinzierl:2003fx,%
Weinzierl:2003ra,Frixione:2004is,Gehrmann-DeRidder:2005hi,Ridder:2005aw,%
Somogyi:2005xz,Weinzierl:2006ij}.
In particular, in \Ref{Somogyi:2006_2}, a subtraction scheme was defined
for computing NNLO corrections to QCD jet cross sections to processes
without coloured partons in the initial state and arbitrary number of
massless particles (coloured or colourless) in the final state. That scheme 
can be summarized as follows.

The NNLO correction to any $m$-jet cross section is a sum of three
contributions, the doubly-real, the one-loop singly-unresolved
real-virtual and the two-loop doubly-virtual terms,
\beq
\tsig{NNLO} =
\int_{m+2}\!\dsig{RR}_{m+2} J_{m+2}
+ \int_{m+1}\!\dsig{RV}_{m+1} J_{m+1}
+ \int_m\!\dsig{VV}_m J_m\:.
\label{eq:sigmaNNLO}
\eeq
Here the notation for the integrals indicate that the double-real
corrections involve the fully-differential cross section
$\dsig{RR}_{m+2}$ of $m+2$ final-state partons, the real-virtual
contribution involves the fully-differential cross section for the
production of $m+1$ final-state partons at one-loop and the double
virtual term is an integral of the fully-differential cross section for
the production of $m$ final-state partons at two-loops over the phase
space of $m$ partons.  The phase spaces are restricted by the
corresponding jet functions $J_n$ that define the physical quantity.

In $d=4$ dimensions the three contributions in \eqn{eq:sigmaNNLO} are
separately divergent, but their sum is finite for infrared-safe
observables. (The requirement of infrared safety implies certain analytic
properties of the jet functions $J_n$ that are spelled out in
\Ref{Somogyi:2005xz}.) As explained in \Ref{Somogyi:2006_2} we first
continue analitically all integrals to $d=4-2\eps$ deminesions and then
rewrite \eqn{eq:sigmaNNLO} as 
\beq
\tsig{NNLO} =
\int_{m+2}\!\dsig{NNLO}_{m+2}
+ \int_{m+1}\!\dsig{NNLO}_{m+1}
+ \int_m\!\dsig{NNLO}_m\,,
\label{eq:sigmaNNLOfin}
\eeq
that is a sum of integrals,
\beeq
\dsig{NNLO}_{m+2} \aand=
\Big\{\dsig{RR}_{m+2} J_{m+2} - \dsiga{RR}{2}_{m+2} J_{m}
     -\Big[\dsiga{RR}{1}_{m+2} J_{m+1} - \dsiga{RR}{12}_{m+2} J_{m}\Big]
\Big\}_{\eps=0}\,,
\label{eq:sigmaNNLOm+2}
\\
\dsig{NNLO}_{m+1} \aand=
\Big\{\Big[\dsig{RV}_{m+1} + \int_1\dsiga{RR}{1}_{m+2}\Big] J_{m+1} 
     -\Big[\dsiga{RV}{1}_{m+1} + \Big(\int_1\dsiga{RR}{1}_{m+2}\Big)
\strut^{{\rm A}_{\scriptscriptstyle 1}}
\Big] J_{m} \Big\}_{\eps=0}\,, \;\;\;\;\;\;
\label{eq:sigmaNNLOm+1}
\eeeq
and
\beq
\dsig{NNLO}_{m} =
\Big\{\dsig{VV}_m + \int_2\Big[\dsiga{RR}{2}_{m+2} - \dsiga{RR}{12}_{m+2}\Big]
     +\int_1\Big[\dsiga{RV}{1}_{m+1} + \Big(\int_1\dsiga{RR}{1}_{m+2}\Big)
\strut^{{\rm A}_{\scriptscriptstyle 1}}
\Big]\Big\}_{\eps=0} J_{m}\,,
\label{eq:sigmaNNLOm}
\eeq
each integrable in four dimesions by construction. Here $\dsiga{RR}{1}_{m+2}$
and $\dsiga{RR}{2}_{m+2}$ are approximate cross sections that
regularize the doubly-real emission cross section in the one- and
two-parton infrared regions of the phase space, respectively. The
double subtraction due to the overlap of these two terms is compensated
by $\dsiga{RR}{12}_{m+2}$. These terms are defined in
\Ref{Somogyi:2006_2} explicitly, where the finiteness of
$\dsig{NNLO}_{m+2}$ is demonstrated also numerically for the case of
$e^+e^- \to $3 jets ($m=3$). In \Ref{Somogyi:2006_1}, we computed the
integral $\int_1\dsiga{RR}{1}_{m+2}$ and showed that the terms in the
first bracket in \eqn{eq:sigmaNNLOm+1} do not contain $\eps$ poles.
Nevertheless, those terms still lead to divergent integrals due to
kinematical singularities in the one-parton unresolved parts of the
phase space. In this paper we define explicitly $\dsiga{RV}{1}_{m+1}$ and
$\Big(\int_1\dsiga{RR}{1}_{m+2}\Big) \strut^{{\rm A}_{\scriptscriptstyle 1}}$,
that regularize the singly-unresolved limits of the real-vitrual cross
section and $\int_1\dsiga{RR}{1}_{m+2}$ in turn. Thus we complete the
presention of all formulae relevant for constructing $\dsig{NNLO}_{m+1}$
explicitly.

\section{Notation}
\label{sec:notation}

\subsection{Matrix elements}
\label{sec:ME}

We consider processes with coloured particles (partons) in the final
states, while the initial-state particles are colourless (tipically
electron-positron annihilation into hadrons).  Any number of additional
non-coloured final-state particles is allowed, too, but they will be
suppressed in the notation.  Resolved partons in the final state are
labelled by $i,k,l,\dots$, the unresolved one is denoted by $r$.

We adopt the colour- and spin-state notation of \Ref{Catani:1996vz}. In
this notation the amplitude for a scattering process involving the
final-state momenta $\mom{}$, $\ket{m}{}{(\mom{})}$, is an abstract
vector in colour and spin space, and its normalization is fixed such
that the squared amplitude summed over colours and spins is
\beq
\label{eq:M2}
|\cM_m|^2 = \bra{m}{}{}\ket{m}{}{}\:.
\eeq
This matrix element has the following formal loop expansion:
\beq
\ket{}{}{} = \ket{}{(0)}{} + \ket{}{(1)}{} + \dots\,,
\label{FormalLoopExpansion}
\eeq
where $\ket{}{(0)}{}$ denotes the tree-level contribution,
$\ket{}{(1)}{}$ is the one-loop contribution and the dots stand for
higher-loop contributions, which are not used in this paper. 

Colour interactions at the QCD vertices are represented by associating
colour charges $\bT_i$ with the emission of a gluon from each
parton $i$.  In the colour-state notation, each vector $\ket{}{}{}$ is
a colour-singlet state, so colour conservation is simply
\beq
\biggl(\sum_j \bT_j \biggr) \,\ket{}{}{} = 0\,,
\label{eq:colourcons}
\eeq
where the sum over $j$ extends over all the external partons of the
state vector $\ket{}{}{}$, and the equation is valid order by order in
the loop expansion of \eqn{FormalLoopExpansion}. 

Using the colour-state notation, we define the two-parton colour-correlated
squared tree amplitudes as
\beeq
|\cM^{(0)}_{(i,k)}(\{p\})|^2 \aand \equiv
\bra{}{(0)}{(\{p\})} \,\bT_i \ldot \bT_k \, \ket{}{(0)}{(\{p\})}
\label{eq:colam2}
\eeeq
and similarly the three-parton colour-correlated squared
tree amplitudes, $|\cM^{(0)}_{(i,k,l)}|^2$ for $i$, $k$ and $l$ being
different, and the doubly two-parton
colour-correlated squared tree amplitudes $|\cM^{(0)}_{(i,k),(j,l)}|^2$:
\beq
|\cM^{(0)}_{(i,k,l)}|^2 \equiv
\sum_{a,b,c} f_{abc} \bra{}{(0)}{} T^a_i T^b_k T^c_l \ket{}{(0)}{}
\label{eq:colam3}
\eeq
and 
\beeq
|\cM^{(0)}_{(i,k)(j,l)}|^2 \aand \equiv
\bra{}{(0)}{}
\,\{\bT_i \cdot \bT_k,\bT_j \cdot \bT_l\} \,
\ket{}{(0)}{}\,,
\label{eq:colam22}
\eeeq
where the anticommutator 
$\{\bT_i \cdot \bT_k,\bT_j \cdot \bT_l\}$
is non-trivial only if $i=j$ or $k=l$, see \eqn{eq:colalg}. We shall also
use the two-parton colour-correlated one-loop amplitude, defined using
an analogous notation:
\beq
2 \Real \bra{}{(0)}{} \ket{}{(1)}{}_{(i,k)}\equiv
2 \Real \bra{}{(0)}{} \,\bT_i \ldot \bT_k \, \ket{}{(1)}{}\,.
\eeq

The colour-charge algebra for the product 
$(\bT_i)^n (\bT_k)^n \equiv \bT_i \ldot \bT_k$ is:
\beq
\bT_i \ldot \bT_k =\bT_k \ldot \bT_i \quad  {\rm if}
\quad i \neq k; \qquad \bT_i^2= C_i\:.
\label{eq:colalg}
\eeq
Here $C_i$ is the quadratic Casimir operator in the representation of
particle $i$ and we have $\CF= \TR(\Nc^2-1)/\Nc= (\Nc^2-1)/(2\Nc)$ in
the fundamental and $\CA=2\,\TR \Nc=\Nc$ in the adjoint representation,
i.e.~we are using the customary normalization $\TR=1/2$.

\subsection{Dimensional regularization, one-loop amplitudes
and renormalization}
\label{sec:dimreg}

We employ conventional dimensional regularization (CDR) in $d=4-2\eps$
space-time dimensions to regulate both the IR and UV divergences,
when quarks (spin-$\hf$ Dirac fermions) possess 2 spin polarizations,
gluons have $d-2$ helicity states and all particle momenta are taken as
$d$-dimensional.

Turning to the renormalization of the amplitudes, let the perturbative
expansion of the scattering amplitude $\cA_m$ in terms
of the bare coupling $\gs \equiv \sqrt{4\pi \asb}$ be
\beq
|\cA_m\ra = \left(\frac{\asb\mu^{2\eps}}{4\pi}\right)^{q/2}
\left[
  |\cA_m^{(0)}\ra
+ \left(\frac{\asb\mu^{2\eps}}{4\pi}\right) |\cA_m^{(1)}\ra
+ \O\left( (\alpha_{\mathrm{s}}^{\mathrm u})^2\right)
\right]\,.
\label{cA}
\eeq
where $q$ is a non-negative integer and $\mu$ is the
dimensional-regularization scale. For the renormalized amplitudes (in
the CDR scheme) we use the notation $|\cM_m\ra$. These are obtained from
the unrenormalized amplitudes by expressing the bare coupling in terms of
the running coupling $\as(\muR2)$ evaluated at the arbitrary
renormalization scale $\muR2$ as
\beq
\asb \mu^{2\eps} = \as(\muR2)\,\muR{2\eps}\,S_\eps^{-1}
\left[1 - \left(\frac{\as(\muR2)}{4\pi}\right)\frac{\beta_0}{\eps}
+ \O(\as^2(\muR2))\right]\,,
\label{couplingrenormalization}
\eeq
where $\beta_0$ is the first coefficient of the $\beta$ function for
$\Nf$ number of light quark flavours,
\beq
\beta_0 = \frac{11}3\CA - \frac43\TR\Nf - \frac23\TR\Ns\,,
\eeq
for QCD $\Ns = 0$.
In \eqn{couplingrenormalization}, $S_\eps$ is the phase space factor
due to the integral over the $(d-3)$-dimensional solid angle, which is
included in the definition of the running coupling in the \msbar\
renormalization scheme,\footnote{The \msbar\ renormalization scheme as
often employed in the literature uses $S_\eps=(4\pi)^\eps \e^{-\eps\gamma_E}$.
It is not difficult to check that the two definitions lead to the same
expressions in a computation at the NLO accuracy. At NNLO these lead to
slightly different bookkeeping of the IR and UV poles at intermediate
steps of the computation, but the physical cross section of
infrared-safe observables is the same.  Our definition leads to
somewhat simpler bookkeeping at the NNLO level.}
\beq
S_\eps = \int\!\frac{\rd^{(d-3)}\Omega}{(2\pi)^{d-3}} =
\frac{(4\pi)^\eps}{\Gamma(1-\eps)}\,.
\eeq
We always consider the running coupling in the \msbar\ scheme defined
with the inclusion of this phase space factor.

The relations between the renormalized amplitudes of
\eqn{FormalLoopExpansion} and the unrenormalized ones are given as
follows:
\beeq
&&
\ket{m}{(0)}{} =
\left(\frac{\as(\muR2)\,\muR{2\eps}}{4\pi} S_\eps^{-1}\right)^{q/2}
|\cA_m^{(0)}\ra
\,,
\label{renA0}
\\ &&
\ket{m}{(1)}{} =
\left(\frac{\as(\muR2)\,\muR{2\eps}}{4\pi} S_\eps^{-1}\right)^{q/2}
\frac{\as(\muR2)}{4\pi} S_\eps^{-1}
\left(\muR{2\eps}\,|\cA_m^{(1)}\ra
 - \frac{q}{2}\frac{\beta_0}{\eps} S_\eps\,|\cA_m^{(0)}\ra\right)
\,.
\label{renA1}
\eeeq

After UV renormalization, the dependence on $\mu$ turns into a
dependence on $\muR{}$, so physical cross sections depend only on
the renormalization scale $\muR{}$.  To avoid a cumbersome notation, we
therefore set $\muR{}=\mu$ in the rest of the paper. Furthermore, after
the IR poles are canceled in an NLO, or NNLO computation, we may set
$\eps = 0$, therefore, the $\muR{2\eps}$ and $S_\eps^{-1}$ factors that
accompany the running coupling in the renormalized amplitude do not 
give any contribution, so we may perform the 
\beq
\left(\frac{\as(\muR2)\,\muR{2\eps}}{4\pi} S_\eps^{-1}\right)^{q/2}
\left(\frac{\as(\muR2)}{4\pi} S_\eps^{-1}\right)^i \to
\left(\frac{\as(\muR2)}{4\pi}\right)^{q/2+i}
\eeq
substitution in \eqns{renA0}{renA1}.

\subsection{Remark on regularization-scheme dependence}
\label{sec:rsdep}

Although the application of conventional dimensional regularization (CDR)
is conceptually clean, the computation of squared matrix elements is much
simpler in other versions of dimensional regularization, most notably in
dimensional reduction (DR). As a result, most of the multiparton QCD
amplitudes $|\cA_m^{(n)}\ra$, both at tree-level ($n=0$) and one-loop
($n=1$), are available in DR. At the level of cross sections however,
the CDR scheme is used traditionally, therefore, the relation between
the two schemes has to be established.

The regularization-scheme (RS) dependence of the matrix elements at
tree level affect only terms of $\Oe{}$, therefore, in computing the
$(m+2)$-parton cross section in \eqn{eq:sigmaNNLOm+2} the RS dependence
is completely harmless, the difference vanishes when we take the
four-dimensional limit. The subtraction terms that regularize the real
emission also depend on the RS. While this dependence does not influence 
$\dsig{NNLO}_{m+2}$, it leads to differences (even in divergent terms)
when the subtraction terms are integrated over the factorized phase
space of the unresolved parton(s). The standard practice in the
literature is to set up the subtraction scheme in CDR and transform the
loop matrix elements to CDR if those were obtained in other schemes, for
instance, in DR.

The RS dependence in the loop amplitudes has in general both ultraviolet
(UV) and infrared (IR) origin. Both have been discussed thoroughly up
to two loops in \Ref{Catani:1998bh}. In the present paper we deal only
with one-loop amplitudes and we summarize the transition rules from DR to
CDR here.

The UV part of the RS dependence is due to the RS dependence of the
renormalization procedure. At the one-loop level it means that
\eqn{renA1} remains valid, with the same expansion parameter, no matter
in which RS the bare amplitudes are computed if we perform the
substitution \cite{Catani:1996vz}
\beq
\beta_0 \to \beta_0 + \eps \tilde{\beta}_0^{\rm RS}
\eeq
in \eqns{couplingrenormalization}{renA1}. By definition in CDR
$\tilde{\beta}_0^{\rm CDR} = 0$. If the bare amplitudes are computed in
the DR scheme, then $\tilde{\beta}_0^{\rm DR} = -\CA/6$.

The IR part of the RS dependence can be decomposed into universal
finite terms and non-universal contributions at $\Oe{}$
\cite{Catani:1996pk}. The finite terms are completely factorized,
\beq
|\cA_{m,{\rm RS}}^{(1){\rm fin}}(\mu^2;\mom{})\ra =
\frac12 \left(\sum_i \ti{\gamma}_i^{\rm RS}\right)
|\cA_{m,{\rm RS}}^{(0)}(\mu^2;\mom{})\ra
+ |F_m^{(1)}(\mu^2;\mom{})\ra
+ \Oe{}\,,
\eeq
while the $\Oe{}$ contributions do not contribute to $\dsig{NNLO}_{m+1}$
in the four-dimensional limit. The transition coefficients that relate
the amplitudes in the RS's depend only on the flavour of the external
partons and were first computed in \Ref{Kunszt:1993sd}. If
$\ti{\gamma}_i^{\rm CDR} = 0$, as always assumed by definition, then
\beq
\ti{\gamma}_q^{\rm DR} = \frac{\CF}{2}\,,\qquad
\ti{\gamma}_g^{\rm DR} = \frac{\CA}{6}\,.
\eeq

\subsection{Cross sections}
\label{sec:xsecs}

In our notation the real-virtual cross section $\dsig{RV}_{m+1}$ is
given by
\beq
\dsig{RV}_{m+1} = \PS{m+1}{(\mom{})}
\:2\Real \la {\cal M}_{m+1}^{(0)}|{\cal M}_{m+1}^{(1)} \ra\,,
\label{eq:dsigRVm+1}
\eeq
where $\PS{m+1}{(\mom{})}$ is the $d$-dimensional phase space for
$m+1$ outgoing particles with momenta
$\mom{} \equiv \{p_1,\dots,p_{m+1}\}$ and total momentum $Q$,
\beq
\label{eq:psf}
\rd\phi_{m+1}(p_1,\dots,p_{m+1};Q) = 
\left[ \,\prod_{i=1}^{m+1} \frac{\rd^{d}p_i}{(2\pi)^{d-1}}
\,\delta_+(p_i^2) \right] 
(2\pi)^d \delta^{(d)}(p_1+\dots+p_{m+1}-Q)\,.
\eeq

The integral of the singly-unresolved approximate cross section for
doubly-real emission over the factorized one-parton phase space was
computed in \Ref{Somogyi:2006_1}
\beq
\int_1\dsiga{RR}{1}_{m+2} = \dsig{R}_{m+1} \otimes \bIt(m,\eps)\,,
\label{eq:I1dsigRRA1_fin2}
\eeq 
where $\dsig{R}_{m+1}$ is the Born-level cross section for the emission
of $m+1$ partons and $\bIt(m,\eps)$ is an operator acting on the colour
space of the $m+1$ final-state partons.  The notation on the right hand
side means that one has to write down the expression for 
$\dsig{R}_{m+1}$ and then replace the Born-level squared matrix element
\beq
\M{m+1}{(0)} = \la {\cal M}_{m+1}^{(0)}|{\cal M}_{m+1}^{(0)} \ra\,,
\eeq
by
\beq
\la {\cal M}_{m+1}^{(0)}|\bIt(m,\eps)|{\cal M}_{m+1}^{(0)} \ra\,.
\label{eq:Mm+1IMm+1}
\eeq
The insertion operator $\bIt(m,\eps)$ differs from the $\bI(\eps)$
operator derived in \Ref{Catani:1996vz} in non-singular terms as $\eps$
tends to zero. Explicitly,
\beeq
&&
\bIt(\mom{};m,\eps) = 
\aeps
\nn\\&&\quad
\times
  \sum_{i} \left[ \IcC{i}{}(y_{iQ};m,\eps) \,\bT_i^2
+ \sum_{k\ne i} \IcS{ik}{}(y_{ik},y_{iQ},y_{kQ};m,\eps)\,\bT_i \bT_k
\right]
\label{eq:bI_def}
\eeeq
where $y_{ik} = s_{ik}/Q^2 \equiv 2p_i\ldot p_k/Q^2$,
$y_{iQ} = 2p_i\ldot Q/Q^2$ and 
with
\beq
\IcC{q}{(0)} = \IcC{qg}{(0)} - \IcCS{(0)}\,,\qquad 
\IcC{g}{(0)} = \frac12 \IcC{gg}{(0)} + \Nf \IcC{q\qb}{(0)} - \IcCS{(0)}\,.
\label{eq:IcCi_def}
\eeq
Explicit expressions for the functions $\IcC{ik}{(0)}(y_{iQ};m,\eps)$,
$\IcS{ik}{(0)}(y_{ik},y_{iQ},y_{kQ};m,\eps)$ and $\IcCS{(0)}(m,\eps)$,
can be found in \Ref{Somogyi:2006_1}, where it was shown that the $\eps$
poles of the one-loop squared matrix element 
$2\Real \la {\cal M}_{m+1}^{(0)}|{\cal M}_{m+1}^{(1)} \ra$
are cancelled exactly by 
$\la {\cal M}_{m+1}^{(0)}|\bIt(\mom{};m,\eps)|{\cal M}_{m+1}^{(0)} \ra$.

%
% Counterterms for the real-virtual cross section
%

\section{Counterterms for the real-virtual cross section}
\label{sec:dsigRVA1}

% Factorization in the singly-unresolved collinear and soft limits

\subsection{Factorization in the collinear and soft limits}

In order to devise the approximate cross section $\dsiga{RV}{1}_{m+1}$,
we have to study the factorization properties of one-loop squared
matrix elements when one parton becomes soft or collinear to another
parton.  The relevant factorization formulae have been computed in
\Refs{Bern:1994zx,Bern:1998sc,Kosower:1999xi,Kosower:1999rx,%
Bern:1999ry,Catani:2000pi}. In our work we use the formulae of
\Ref{Bern:1999ry} for collinear parton splitting and those in
\Ref{Catani:2000pi} for soft gluon emission.  However, the notation in
those papers is not convenient for writing factorization formulae which
avoid double counting in the soft-collinear limit, therefore, we
present new formulae here. 

\subsubsection{Collinear limit}

We define the collinear limit of two final-state momenta $p_i$ and $p_r$
with the help of an auxiliary light-like vector $n_{ir}^\mu$
($n_{ir}^2=0$) using the usual Sudakov parametrization,
\beq
p_i^\mu = z_i p_{ir}^\mu - \kT{r}^\mu 
- \frac{\kT{r}^2}{z_i}\frac{n_{ir}^\mu}{2p_{ir}n_{ir}}\:, \quad
p_r^\mu = z_r p_{ir}^\mu + \kT{r}^\mu 
- \frac{\kT{r}^2}{z_r}\frac{n_{ir}^\mu}{2p_{ir}n_{ir}}\:, \quad
\label{eq:sudakov}
\eeq
where $p_{ir}^\mu$ is a light-like momentum that points towards the
collinear  direction and $\kT{r}$ is the momentum component that is
orthogonal to both $p_{ir}$ and $n_{ir}$ ($p_{ir}\cdot \kT{r}=
n_{ir}\cdot \kT{r}=0$). Momentum conservation requires that $z_i+z_r=1$.
The two-particle invariant mass of the collinear partons is 
\beq
\label{eq:scdouble}
s_{ir} = -\frac{\kT{r}^2}{z_i z_r}\:.
\eeq
The collinear limit is defined by the uniform rescaling 
\beq
\kT{r}\to \lambda \kT{r}\,,
\label{eq:kTscaling}
\eeq
and taking the limit $\lambda \to 0$, when the one-loop squared 
matrix element of an $(m+1)$-parton process has the following 
asymptotic form \cite{Bern:1999ry}:%
\footnote{Since we deal with final-state singularities only, we have
$s_{ir} > 0$ and we can write the usual factor
$(-\mu^2/s_{ir})^\eps$ as $(\mu^2/s_{ir})^\eps \cos (\pi\eps)$.}
\beeq
&&
2\Real \bra{m+1}{(0)}{(p_i,p_r,\ldots)}\ket{m+1}{(1)}{(p_i,p_r,\ldots)}
\simeq
\nn\\ &&\quad
8\pi\as\mu^{2\eps}
\frac{1}{s_{ir}}
\Bigg[2 \Real \bra{m}{(0)}{(p_{ir},\ldots)}
      \hP_{f_i f_r}^{(0)} \ket{m}{(1)}{(p_{ir},\ldots)}
\nn\\ &&\qquad\qquad\qquad
+\,8\pi\as\,c_\Gamma\left(\frac{\mu^2}{s_{ir}}\right)^\eps \cos(\pi\eps)
\,\bra{m}{(0)}{(p_{ir},\ldots)}\hP_{f_i f_r}^{(1)}\ket{m}{(0)}{(p_{ir},\ldots)}
\Bigg]
\label{eq:RVcollfact}
\eeeq
where
\beq
c_{\Gamma}=\frac{1}{(4\pi)^{2-\eps}}
\frac{\Gamma(1+\eps)\Gamma^2(1-\eps)}{\Gamma(1-2\eps)}\,.
\eeq
The meaning of the $\simeq$ sign in \eqn{eq:RVcollfact} is that we have
neglected subleading terms (in this case those that are less singular
than $1/\lambda^2$).  In order to simplify further discussion, following
the notation of \Ref{Somogyi:2005xz}, we introduce a symbolic operator
$\bC{ir}$ that performs the action of taking the collinear limit of the
one-loop squared matrix element, keeping the leading singular term:
\beeq
&&
\bC{ir}
\,2\Real \bra{m+1}{(0)}{(p_i,p_r,\ldots)}\ket{m+1}{(1)}{(p_i,p_r,\ldots)} =
\nn\\ &&\quad
8\pi\as\mu^{2\eps}
\frac{1}{s_{ir}}
\Bigg[2 \Real \bra{m}{(0)}{(p_{ir},\ldots)}
      \hP_{f_i f_r}^{(0)} \ket{m}{(1)}{(p_{ir},\ldots)}
\nn\\ &&\qquad\qquad\qquad
+\,8\pi\as\,c_\Gamma\left(\frac{\mu^2}{s_{ir}}\right)^\eps \cos(\pi\eps)
\,\bra{m}{(0)}{(p_{ir},\ldots)}\hP_{f_i f_r}^{(1)}\ket{m}{(0)}{(p_{ir},\ldots)}
\Bigg]
\,.
\label{eq:RVcollfactnew}
\eeeq

The $m$-parton matrix elements on the right-hand side of
\eqn{eq:RVcollfact} are obtained from the $(m+1)$-parton matrix
elements by removing partons $i$ and $r$ and replacing them with a
single parton denoted by $ir$. The parton $ir$ carries the quantum
numbers of the pair $i + r$ in the collinear limit: its momentum is
$p_{ir}^{\mu}$ and its other quantum numbers (flavour, colour) are
obtained according to the following rule: anything + gluon gives
anything and quark + antiquark gives gluon.  The kernels $\hP_{f_i
f_r}^{(0)}$ and $\hP_{f_i f_r}^{(1)}$ are the $d$-dimensional
Altarelli-Parisi splitting functions and their one-loop corrections,
which depend on the momentum fractions of the decay products and on the
relative transverse momentum of the pair. For the sake of simplicity,
we label the momentum fractions belonging to a certain parton flavour
with the corresponding label of the squared matrix element, $z_{f_i}=z_i$.
In the case of splitting into a pair, only one momentum fraction is
independent, yet, we find it more convenient to keep the functional
dependence on both $z_i$ and $z_r$. Depending on the $f_i$ flavours
of the splitting products the explicit functional forms are%
\footnote{We remind the reader that the formulae are valid in the CDR scheme.}
\beeq
\la\mu|\hP_{g_ig_r}^{(0)}(z_i,z_r,\kTm{\mu};\eps)|\nu\ra \aand=
2\CA\left[-g^{\mu\nu}\left(\frac{z_i}{z_r}+\frac{z_r}{z_i}\right)-2(1-\eps)
z_i z_r \frac{\kTm{\mu}\kTm{\nu}}{\kTm{2}}\right]\,,
\label{P0gg}
\\
\la\mu|\hP_{\qb_i q_r}^{(0)}(z_i,z_r,\kTm{\mu};\eps)|\nu\ra \aand=
\TR\left[
-g^{\mu\nu}+4z_i z_r \frac{\kTm{\mu}\kTm{\nu}}{\kTm{2}}
\right]\,,
\label{P0qq}
\\
\la r|\hP_{q_ig_r}^{(0)}(z_i,z_r;\eps)|s\ra \aand=
\delta_{rs}\CF\left[\frac{1+z_i^2}{z_r}-\eps z_r \right]
\equiv \delta_{rs} P^{(0)}_{q_ig_r}(z_i,z_r;\eps)\,,
\label{P0qg}
\eeeq
where in the last equation we introduced our notation for the
spin-averaged splitting function,
\beq
P_{f_i f_r}(z_i,z_r;\eps) \equiv \la\hP_{f_i f_r}(z_i,z_r,\kTm{\mu};\eps)\ra\:.
\label{eq:Pav}
\eeq
The one-loop kernels are
\beeq
\la\mu|\hP_{g_i g_r}^{(1)}(z_i,z_r,\kTm{\mu};\eps)|\nu\ra \aand=
r_{\S,\rm ren}^{g_i g_r}(z_i,z_r;\eps)
\la\mu|\hP_{g_i g_r}^{(0)}(z_i,z_r,\kTm{\mu};\eps)|\nu\ra
\nn\\ \aand-
\,4\,\CA\,r_\NS^{g_i g_r}\Big[1-2\eps z_i z_r \Big]
\frac{\kTm{\mu}\kTm{\nu}}{\kTm{2}}\,,
\label{P1gg}
\\
\la\mu|\hP_{\qb_i q_r}^{(1)}(z_i,z_r,\kTm{\mu};\eps)|\nu\ra \aand=
r_{\S,\rm ren}^{\qb_i q_r}(z_i,z_r)
\la\mu|\hP_{\qb_i q_r}^{(0)}(z_i,z_r,\kTm{\mu};\eps)|\nu\ra\,,
\label{P1qq}
\\
\la r|\hP_{q_i g_r}^{(1)}(z_i,z_r;\eps)|s \ra \aand=
r_{\S,\rm ren}^{q_i g_r}(z_i,z_r) \la r|\hP_{q_i g_r}^{(0)}(z_i,z_r;\eps)|s\ra
+ \delta_{rs}\,\CF\,r_\NS^{q_i g_r}\Big[1-\eps z_r\Big]\,.
\label{P1qg}
\eeeq
The $r_{\S,\rm ren}(z_i,z_r;\eps)$ singular factors are expressed in
terms of corresponding unrenormalized $r_\S(z_i,z_r;\eps)$ factors. The
relation between the two forms is given by the equation:
\beeq
&&
r_{\S,\rm ren}(z_i,z_r;\eps)=
r_\S(z_i,z_r;\eps)
- \frac{\beta_0}{2\eps}\,\frac{S_\eps}{(4\pi)^2 c_\Gamma}
\,\left[\left(\frac{\mu^2}{s_{ir}}\right)^{\eps}\cos(\pi\eps)\right]^{-1}
\,.
\label{rSUV}
\eeeq
The unrenormalized $r_\S^{f_i f_r}(z_i,z_r)$ factors and the
$r_\NS^{f_i f_r}$ factors can be trivially obtained  from the
${\cal D}_{f_{ir} \to f_i f_r}^{\mu,\,{\rm 1-loop}}$ functions, that were
computed in \Ref{Bern:1999ry}. In the case of gluon splitting those
functions are symmetric under the exchange of $z_i$ and $z_r$. To make
this symmetry manifest, we have re-cast the original expression of
\Ref{Bern:1999ry} for the gluon splitting into a $q\qb$ pair into an
equivalent form which exhibits the $z_i \leftrightarrow z_r$ symmetry,
\beeq
r_\S^{\qb_i q_r}(z_i,z_r) \aand=
\frac{1}{\eps^2} (\CA - 2\CF)
+ \frac{\CA}{\eps^2}
\sum_{m=1}^{\infty} \eps^m
\left[\Li_m \left(\frac{-z_i}{z_r}\right)
    + \Li_m \left(\frac{-z_r}{z_i}\right) \right]
\nn \\ &&
+ \frac{1}{1 - 2\eps}\Bigg[
  \frac{1}{\eps}
  \left(\frac{11}{3}\CA -\frac{4\TR}{3}\Nf-\frac{2\TR}{3}\Ns - 3\CF\right)
\nn \\ &&\qquad\qquad
+ \CA - 2\CF + \frac{\CA + 4\TR(\Nf - \Ns)}{3(3-2\eps)}
\Bigg] \,.
\label{rSqq}
\eeeq
In the case of gluon splitting into two gluons, the same symmetry is
valid, which, however, we choose not to make manifest. Instead, we use a
form, where the polylogarithms are regular in the $z_r\to 0$ limit
(which will be convenient when we compute the soft limit of this
expression in \sect{sec:RVmatchingcollsoft}, see \eqn{eq:SrCirRVnew}),
\beeq
r_\S^{g_i g_r}(z_i,z_r) \aand=
- \frac{\CA}{\eps^2} 
\left[
\left(\frac{z_i}{z_r}\right)^\eps \frac{\pi \eps}{\sin(\pi \eps)}
- \sum_{m=1}^{\infty} 2\eps^{2m-1} \Li_{2m-1} \left(\frac{-z_r}{z_i}\right)
\right] \,,
\label{rSgg}
\eeeq
and similarly in the case of quark splitting,
\beq
r_\S^{q_i g_r}(z_i,z_r) = - \frac{1}{\eps^2} 
\left[\CA \left(\frac{z_i}{z_r}\right)^{\eps} \frac{\pi\eps}{\sin(\pi\eps)}
+ \sum_{m=1}^{\infty}\eps^m\Big[(1 + (-1)^m)\CA - 2\CF\Big]
  \Li_m \left(\frac{-z_r}{z_i}\right)\right]
\,.
\label{rSqg}
\eeq
\eqn{rSgg} also shows that polylogarithms with even subscripts do not
appear in the $\eps$ expansion of the $r_\S^{g_i g_r}$ singular function for gluon 
splitting.

The $r_\NS$ non-singular factors do not depend on the momentum fractions,
\beq
r_\NS^{g_i g_r} =
\frac{\CA(1-\eps)-2\TR (\Nf-\Ns)}{(1-2\eps)(2-2\eps)(3-2\eps)}\,,
\qquad
r_\NS^{q_i g_r} =
\frac{\CA-\CF}{1-2\eps}\,.
\label{rNSqg}
\eeq
In \eqns{rSqq}{rNSqg} $\Nf$ and $\Ns$ denote the number of fundamental
fermions and scalars that can circulate in the loops. The case of QCD
is obtained by setting $\Ns = 0$.

The gluon-gluon and quark-antiquark splittings are symmetric in the 
momentum fractions of the two decay products (even though this is not
manifest for gluon-gluon splitting), while the quark-gluon
splitting is not. Nevertheless, we do not distinguish the flavour
kernels $\hP_{qg}$ and $\hP_{gq}$. The ordering of the flavour indices
and arguments of the Altarelli-Parisi kernels has no meaning in our
notation, i.e.,
\beq
\hP_{f_i f_r}(z_i,z_r;\eps) = \hP_{f_r f_i}(z_r,z_i;\eps)\,.
\label{eq:PirPri}
\eeq
Thus, it is sufficient to record the kernel belonging to one ordering.
We keep this convention throughout.

\subsubsection{Soft limit}

The soft limit is defined by parametrizing the soft momentum as
$p_r^\mu = \lambda q_r^\mu$ and letting $\lambda \to 0$ at fixed
$q_r^\mu$. Neglecting terms less singular than $1/\lambda^2$, it was
found \cite{Catani:2000pi} that 
\beeq
&&
2\Real \bra{m+1}{(0)}{(p_r,\ldots)}\ket{m+1}{(1)}{(p_r,\ldots)}
\simeq
\nn\\ &&\qquad
-8\pi\as\mu^{2\eps}
\sum_{i}\,\sum_{k\ne i}\,\frac12 \calS_{ik}(r)
\Bigg\{
2 \Real \bra{m}{(0)}{(\ldots)}\bT_i \bT_k \ket{m}{(1)}{(\ldots)}
\nn \\ && \qquad
-8\pi\as c_\Gamma\Bigg[\left(\CA
\frac{1}{\eps^2} \frac{\pi\eps}{\sin({\pi\eps})}
\left(\frac12 \mu^2 \calS_{ik}(r)\right)^\eps \cos(\pi\eps)
+ \frac{\beta_0}{2\eps}\frac{S_\eps}{(4\pi)^2 c_\Gamma}\right)
|\cM_{m;(i,k)}^{(0)}(\ldots)|^2
\nn \\ && \qquad\qquad\qquad
-2\frac{\pi}{\eps} 
\sum_{l\ne i,k}
\left(\frac12 \mu^2 \calS_{kl}(r) \right)^{\eps}
|\cM_{m;(i,k,l)}^{(0)}(\ldots)|^2\Bigg]\Bigg\}
\label{eq:RVsoftfact}
\eeeq
if $r$ is a gluon.  Similarly to the $\bC{ir}$ operator of taking the
collinear limit, following \Ref{Somogyi:2005xz} we introduce another
symbolic operator $\bS{r}$ that performs the action of taking the soft
limit of the squared matrix element, keeping the leading singular terms.
With this notation
$\bS{r}\,2\Real \bra{m+1}{(0)}{(p_r,\ldots)}\ket{m+1}{(1)}{(p_r,\ldots)}$
is equal to the right hand side of \eqn{eq:RVsoftfact} if $r$ is a gluon
and
$\bS{r}\,2\Real \bra{m+1}{(0)}{(p_r,\ldots)}\ket{m+1}{(1)}{(p_r,\ldots)} = 0$
if $r$ is a quark.

In \eqn{eq:RVsoftfact} the $m$-parton matrix element on the right-hand
side is obtained from the $(m+1)$-parton matrix element on the
left-hand side by simply removing the soft parton. The eikonal factor is 
\beq
\calS_{ik}(r) = \frac{2 s_{ik}}{s_{ir} s_{rk}}\,.
\label{eq:Sikr}
\eeq
Note that \eqn{eq:RVsoftfact} is valid only for the case of final-state
partons.  The general case can be found in \cite{Catani:2000pi}.  

\subsubsection{Matching the collinear and soft limits}
\label{sec:RVmatchingcollsoft}

If we want to regularize the squared matrix elements in all
singly-unresolved regions of the phase space then we have to subtract
all possible collinear and soft limits, i.e. subtract the sum
\beq
\sum_r\left(\sum_{i\ne r}\frac12 \bC{ir} + \bS{r}\right)
2\Real \bra{m+1}{(0)}{(p_i,p_r,\ldots)}\ket{m+1}{(1)}{(p_i,p_r,\ldots)}\,,
\label{eq:Acandidate}
\eeq
where the $1/2$ symmetry factor because in the summation each collinear
configuration is taken into account twice. Subtracting \eqn{eq:Acandidate}
we perform a double subtraction in some regions of the phase space
where the soft and collinear limits overlap. In order to compensate
for the double subtraction, we need to find the collinear limit of
the right hand side of \eqn{eq:RVsoftfact} when gluon $r$ becomes
simultaneously collinear to parton $i$. In deriving this limit we use
that in the collinear limits 
(i)
the factors multiplying two-parton colour-correlated squared matrix
elements are independent of $k$, therefore using colour conservation
(\eqn{eq:colourcons}) we can perform the summation over $k$;
(ii)
the factors multiplying the three-parton colour-correlated squared
matrix element are symmetric in $k$ and $l$ while
$|\cM_{m;(i,k,l)}^{(0)}(p_i,\ldots)|^2$ is antisymmetric, 
thus the sum of those terms is zero. Finally we have
\beeq
&&
\bC{ir}\bS{r}
\,2\Real \bra{m+1}{(0)}{(p_i,p_r,\ldots)}\ket{m+1}{(1)}{(p_i,p_r,\ldots)} =
8\pi\as\mu^{2\eps} \frac{2}{s_{ir}} \frac{z_i}{z_r}\bT_i^2
\nn\\ &&\quad
\times
\Bigg[
2 \Real \bra{m}{(0)}{(p_i,\ldots)} \ket{m}{(1)}{(p_i,\ldots)}
\nn \\ && \quad
- 8\pi\as c_\Gamma \left(\CA
\frac{1}{\eps^2} \frac{\pi\eps}{\sin({\pi\eps})}
\left(\frac{\mu^2}{s_{ir}} \frac{z_i}{z_r}\right)^\eps \cos(\pi\eps)
+ \frac{\beta_0}{2\eps}\frac{S_\eps}{(4\pi)^2 c_\Gamma}\right)
|\cM_{m}^{(0)}(p_i,\ldots)|^2\Bigg]
\,.\quad~
\label{eq:CirSrRVnew}
\eeeq
Similarly, the soft limit of \eqn{eq:RVcollfactnew} when $r$ is a gluon
and $z_r \to 0$ is
\beeq
&&
\bS{r}\bC{ir}
\,2\Real \bra{m+1}{(0)}{(p_i,p_r,\ldots)}\ket{m+1}{(1)}{(p_i,p_r,\ldots)} =
8\pi\as\mu^{2\eps} \frac{2}{s_{ir}} \frac{1}{z_r}\bT_i^2
\nn \\ &&\quad
\times
\Bigg[
2 \Real \bra{m}{(0)}{(p_i,\ldots)} \ket{m}{(1)}{(p_i,\ldots)}
\nn \\ && \quad
- 8\pi\as c_\Gamma \left(\CA
\frac{1}{\eps^2} \frac{\pi\eps}{\sin({\pi\eps})}
\left(\frac{\mu^2}{s_{ir}} \frac{1}{z_r}\right)^\eps \cos(\pi\eps)
+ \frac{\beta_0}{2\eps}\frac{S_\eps}{(4\pi)^2 c_\Gamma}\right)
|\cM_{m}^{(0)}(p_i,\ldots)|^2\Bigg]
\,.\quad~
\label{eq:SrCirRVnew}
\eeeq
\eqns{eq:CirSrRVnew}{eq:SrCirRVnew} differ by the term $z_i = 1 - z_r$ 
in the numerator of \eqn{eq:CirSrRVnew}, which is subleading if $r$ is
soft. Therefore, \eqn{eq:CirSrRVnew} can be used to account for the
double subtraction: it cancels the soft subtraction in the collinear
limit by construction,
\beq
\bC{ir}(\bS{r}-\bCS{ir}{r}) 2\Real \bra{m+1}{(0)}{}\ket{m+1}{(1)}{} = 0\,,
\label{eq:CirSr-CSir}
\eeq
and the $\bC{ir}-\bCS{ir}{r}$ difference is subleading in the soft limit,
\beq
\bS{r}(\bC{ir}-\bCS{ir}{r}) 2\Real \bra{m+1}{(0)}{}\ket{m+1}{(1)}{} = 0\,.
\label{eq:SrCir-CSir}
\eeq
Accordingly, in order to remove the double subtraction from
\eqn{eq:Acandidate}, we have to add terms like that in
\eqn{eq:CirSrRVnew}.  That amounts to always take the collinear limit
of the soft factorization formula rather than the reverse (like terms
in \eqn{eq:SrCirRVnew}).  Thus the candidate for a subtraction term for
regularizing the squared matrix element in all singly-unresolved limits is
\beeq
&&
\bA{1}\,2\Real \bra{m+1}{(0)}{}\ket{m+1}{(1)}{} = 
\label{eq:RV_A1}
\\ &&\qquad
\nn
\sum_r \left[\sum_{i\ne r} \frac12 \bC{ir}
+\left(\bS{r}-\sum_{i\ne r} \bCS{ir}{r}\right)\right]
2\Real \bra{m+1}{(0)}{(p_i,p_r,\ldots)}\ket{m+1}{(1)}{(p_i,p_r,\ldots)}\,.
\eeeq
Note that the cancellation of the collinear terms in the soft limit
actually requires the symmetry factor multiplying the collinear term, but
not the collinear-soft one. This form of the $\bA{1}$ operator coincides
with that derived in \Ref{Somogyi:2005xz} for separating the
singly-unresolved kinematical singularities of the squared matrix
element at tree-level and is completely universal.

\subsection{Counterterms}
\label{sec:RVcounterterms}

The expression given in \eqn{eq:RV_A1} is defined only in the strict soft
and/or collinear limits. In order to define true countertems, we have to
extend it over the whole phase space. This extension requires an exact
factorization of the $m+1$ parton phase space into an $m$ parton phase
space times the phase space measure of the unresolved parton,
\beq
\PS{m+1}{(\mom{})} = \PS{m}{(\momt{})}\,[\rd p_1]\,,
\label{eq:PS_fact}
\eeq
where we introduced the compact notations $\mom{}\equiv\{p_1,\ldots,p_{m+1}\}$
and $\momt{}\equiv\{\ti{p}_1,\ldots,\ti{p}_m\}$.
Then the subtraction term that regularizes the kinematical singularities
of the real-virtual cross section can symbolically be written as
\beq
\dsiga{RV}{1}_{m+1} = 
\PS{m}{}[\rd p_1]\,\bcA{1}{(1)}
\,2\Real \bra{m+1}{(0)}{}\ket{m+1}{(1)}{}\,.
\label{eq:dsigRVA1}
\eeq
where we decompose the subtraction term as follows,
\beeq
&&
\bcA{1}{(1)} \,2\Real \bra{m+1}{(0)}{}\ket{m+1}{(1)}{} =
\nn\\&&\qquad
= \sum_{r} \left[\sum_{i\ne r} \frac{1}{2} \cC{ir}{(0,1)}(\mom{})
+ \left(\cS{r}{(0,1)}(\mom{}) - \sum_{i\ne r} \cCS{ir}{r}{(0,1)}(\mom{})\right) 
\right]
\nn\\&&\qquad
+ \sum_{r} \left[\sum_{i\ne r} \frac{1}{2} \cC{ir}{(1,0)}(\mom{})
+ \left(\cS{r}{(1,0)}(\mom{}) - \sum_{i\ne r} \cCS{ir}{r}{(1,0)}(\mom{})\right) 
\right]\,.
\label{eq:A11M0M1}
\eeeq
All terms above are functions of the original $m+1$ momenta that enter
the one-loop squared matrix element. The last terms in each line on
the right hand side do not refer to the collinear limit of anything, but
denote functions of the original momenta for which the notation
inherits the operator structure of taking the various limits, but
otherwise has nothing to do with taking limits.

We now turn to the definition of each term in \eqn{eq:A11M0M1}. Each
term will have the structure that a singular function (Altarelli-Parisi
kernel or eikonal factor) is sandwiched between amplitudes. Both the
singular functions and the squared matrix elements have their own loop
expansions. The double superscipt on the subtraction terms refers to the
number of loops in these loop expansions, the first one in the loop
expansion of the singular factor while the second one in the expansion of
the squared matrix element.

\subsubsection{Collinear counterterms}

The collinear counterterms are
\beeq
&&
\cC{ir}{(0,1)}(\mom{}) =
8\pi\as\mu^{2\eps}\frac{1}{s_{ir}}
\nn\\&&\qquad
\times
2\Real\bra{m}{(0)}{(\momt{(ir)})}
\hP_{f_i f_r}^{(0)}(\tzz{i}{r},\tzz{r}{i},\kTt{i,r};\eps)
\ket{m}{(1)}{(\momt{(ir)})}\,,
\label{eq:Cir01}
\\[2mm] &&
\cC{ir}{(1,0)}(\mom{}) =
(8\pi\as\mu^{2\eps})^2\frac{1}{s_{ir}^{1+\eps}} c_{\Gamma}\cos(\pi\eps)
\nn\\&&\qquad
\times
\bra{m}{(0)}{(\momt{(ir)})}
\hP_{f_i f_r}^{(1)}(\tzz{i}{r},\tzz{r}{i},\kTt{i,r};\eps)
\ket{m}{(0)}{(\momt{(ir)})}\,.
\label{eq:Cir10}
\eeeq
The momentum fractions $\tzz{i}{r}$ and $\tzz{r}{i}$ are
\beq
\tzz{i}{r} = \frac{y_{iQ}}{y_{(ir)Q}}
\qquad\mbox{and}\qquad
\tzz{r}{i} = \frac{y_{rQ}}{y_{(ir)Q}}\,,
\label{eq:zt2}
\eeq
while the transverse momentum $\kTt{i,r}$ is
\beq
\kTt{i,r}^{\mu} = 
\zeta_{i,r} p_r^{\mu} - \zeta_{r,i} p_i^{\mu} + \zeta_{ir} \ti{p}_{ir}^{\mu}
\,,\qquad
\zeta_{i,r} = \tzz{i}{r}-\frac{y_{ir}}{\alpha_{ir}y_{(ir)Q}}
\,,\quad
\zeta_{r,i} =  \tzz{r}{i}-\frac{y_{ir}}{\alpha_{ir}y_{(ir)Q}}
\,.
\label{eq:kTtir}
\eeq
We used the abbreviations $y_{ir}= s_{ir}/Q^2 \equiv 2p_i\cdot p_r/Q^2$,
$y_{(ir)Q} = y_{iQ} + y_{rQ}$ with $y_{iQ}=2p_i\cdot Q/Q^2$,
$y_{rQ}=2p_r\cdot Q/Q^2$ and $Q^\mu$ is the total four-momentum of the
incoming electron and positron, while $\ti{p}_{ir}^{\mu}$ and
$\alpha_{ir}$ are defined below in \eqns{eq:PS_Cir}{eq:alphair}
respectively.  This choice for the transverse momentum is exactly
perpendicular to the parent momentum $\ti{p}_{ir}^{\mu}$ and ensures
that in the collinear limit $p_i^\mu || p_r^\mu$, the square of
$\kTt{i,r}^{\mu}$ behaves as
\beq
\kTt{i,r}^2 \simeq - s_{ir}\tzz{r}{i} \tzz{i}{r}
\,,
\label{eq:kTir2}
\eeq
as required (independently of $\zeta_{ir}$). In our computation the
longitudinal component, proportional to $\zeta_{ir}$, does not
contribute due to gauge invariance of the matrix elements, therefore,
we may choose $\zeta_{ir} = 0$. The $m$ momenta
$\momt{(ir)} \equiv \{\ti{p}_1,\ldots,\ti{p}_{ir},\ldots,\ti{p}_{m+1}\}$ 
entering the matrix elements on the right hand side of
\eqns{eq:Cir01}{eq:Cir10} are
\beq
\ti{p}_{ir}^{\mu} = \frac{1}{1-\alpha_{ir}}(p_i^{\mu} + p_r^{\mu} - \alpha_{ir} Q^{\mu})\,,
\qquad
\ti{p}_n^{\mu} = \frac{1}{1-\alpha_{ir}} p_n^{\mu}\,,
\qquad n\ne i,r\,,
\label{eq:PS_Cir}
\eeq
where
\beq
\alpha_{ir} =
\frac12\Big[y_{(ir)Q}-\sqrt{y_{(ir)Q}^2 - 4y_{ir}}\;\Big]\,.
\label{eq:alphair}
\eeq

This momentum mapping leads to an exact factorization of the phase space
in the form of \eqn{eq:PS_fact}. The explicit expression for
$[\rd p_{1}]$ reads
\beq
[\rd p_{1;m}^{(ir)}(p_r,\ti{p}_{ir};Q)] =
\Jac{ir}{1}(p_r,\ti{p}_{ir};Q)
\,\frac{\rd^d p_r}{(2\pi)^{d-1}}\delta_{+}(p_r^2)\,,
\label{eq:dp_Cir}
\eeq
where the Jacobian is
\beq
\Jac{ir}{1}(p_r,\ti{p}_{ir};Q) =
y_{\wti{ir}Q}\,\frac{(1-\alpha_{ir})^{(m-1)(d-2)-1}\,\Theta(1-\alpha_{ir})}
{2 (1 - y_{\wti{ir}Q}) \alpha_{ir} + y_{r\wti{ir}}+y_{\wti{ir}Q}-y_{rQ}}
\,.
\label{eq:Jac_Cir}
\eeq
In this equation $\alpha_{ir}$ has to be expressed as a function of the
variable $\ti{p}_{ir}$,
\beq
\alpha_{ir}=
\frac{\sqrt{(y_{r\wti{ir}}+y_{\wti{ir}Q}-y_{rQ})^2
           + 4y_{r\wti{ir}}(1-y_{\wti{ir}Q})}
-(y_{r\wti{ir}}+y_{\wti{ir}Q}-y_{rQ})}{2(1-y_{\wti{ir}Q})}\,.
\label{eq:alpha_ir_new}
\eeq

It is straightforward to compute the $\eps$ expansion of the collinear
counterterms that we shall use later. We decompose the expansion into
singular terms, finite contributions and terms that vanish as $\eps \to 0$,
\beq
\cC{ir}{(0,1)}(\mom{}) =
\pole\cC{ir}{(0,1)}(\mom{}) +
\finite\cC{ir}{(0,1)}(\mom{}) + \Oe{}\,.
\label{eq:pole+fin}
\eeq
The pole parts can be written in the following unified form:
\beeq
&&
\pole\cC{ir}{(0,1)}(\mom{}) =
- 8\pi\as\mu^{2\eps}\frac{1}{s_{ir}}
\nn\\&&\qquad
\times
\bra{m}{(0)}{(\momt{(ir)})}
\,\bI(\momt{(ir)};\eps)
\,\hP_{f_i f_r}^{(0)}(\tzz{i}{r},\tzz{r}{i},\kTt{i,r};\eps)
\,\ket{m}{(0)}{(\momt{(ir)})}
\,,
\label{eq:polesCir01}
\eeeq
where%
\footnote{Note that 
$\bIt(m,\eps) = \bI(\eps) + \Oe{0}$ independently of $m$.}
\beq
\bI(\momt{(ir)};\eps) = \aeps
\sum_i\left( \bT_i^2 \frac{1}{\eps^2} + \gamma_i \frac{1}{\eps}
+ \sum_{k\ne i} \bT_i \bT_k \frac{1}{\eps} \ln y_{\ti{i}\ti{k}}\right)\,
\label{eq:bI}
\eeq
with the usual flavour constants
\beq
\gamma_q = \frac32 \CF\,,\qquad \gamma_g = \frac{\beta_0}{2}\,.
\label{eq:gammadef}
\eeq
The poles of the $\cC{ir}{(1,0)}(\mom{})$ counterterm can be written as
\beeq
&&
\pole\cC{ir}{(1,0)}(\mom{}) =
-8\pi\as\mu^{2\eps}\frac{1}{s_{ir}} \frac{\as}{2\pi} S_\eps
\nn\\&&\qquad
\times
\Bigg[
(\bT_i^2 + \bT_r^2 - \bT_{ir}^2)
\left(\frac{1}{\eps^2}-\frac{1}{\eps} \ln y_{ir}\right)
+ \frac{1}{\eps} \Big( \gamma_i + \gamma_r - \gamma_{ir} \Big) 
\nn\\&&\qquad \qquad
- \frac{1}{\eps}
  \Big((\bT_i^2 - \bT_r^2 + \bT_{ir}^2) \ln \tzz{i}{r}
      +(\bT_r^2 - \bT_i^2 + \bT_{ir}^2) \ln \tzz{r}{i} \Big) 
\Bigg]
\nn\\&&\qquad
\times
\bra{m}{(0)}{(\momt{(ir)})}
\hP_{f_i f_r}^{(0)}(\tzz{i}{r},\tzz{r}{i},\kTt{i,r};\eps)
\ket{m}{(0)}{(\momt{(ir)})}
\,,
\label{eq:polesCir10}
\eeeq
where $\bT_{ir} = \bT_i + \bT_r$. After the cancellation of the poles
is demonstrated (see \sect{sec:A1A1counterterms}), in a computer code
one uses the finite parts of the counterterms. We collect all such
finite contributions of the counterterms in \app{app:finiteparts}.

\subsubsection{Soft-type counterterms}

We call the soft and soft-collinear counterterms soft-type terms
because they all use the momentum mapping appropriate to the soft
countertem. We define 
\beeq
\cS{r_g}{(0,1)}(\mom{}) \aand= 
-8\pi\as\mu^{2\eps}\sum_{i}\sum_{k\ne i} \frac12 \calS_{ik}(r)
2\Real \bra{m}{(0)}{(\momt{(r)})} \bT_{i} \bT_{k} \ket{m}{(1)}{(\momt{(r)})}\,,
\label{eq:Sr01}
\\[3mm]
\cCS{ir_g}{r_g}{(0,1)}(\mom{}) \aand= 
8\pi\as\mu^{2\eps} \frac{1}{s_{ir}}\frac{2\tzz{i}{r}}{\tzz{r}{i}}\,\bT_i^2\,
2\Real \bra{m}{(0)}{(\momt{(r)})} \ket{m}{(1)}{(\momt{(r)})}\,,
\label{eq:CirSr01}
\\[3mm]
\cS{r_g}{(1,0)}(\mom{}) \aand=
(8\pi\as\mu^{2\eps})^2 c_{\Gamma}\sum_{i}\sum_{k\ne i} \frac12 \calS_{ik}(r)
\nn\\&&\times
\Bigg[\left(\CA
\frac{1}{\eps^2} \frac{\pi\eps}{\sin({\pi\eps})}
\left(\frac12 \calS_{ik}(r)\right)^\eps \cos(\pi\eps)
+ \frac{\beta_0}{2\eps}\frac{\mu^{-2\eps}S_\eps}{(4\pi)^2 c_\Gamma}\right)
|\cM_{m;(i,k)}^{(0)}(\momt{(r)})|^2
\nn \\ && \quad
-2\frac{\pi}{\eps} 
\sum_{l\ne i,k}
\left(\frac12 \calS_{kl}(r) \right)^{\eps}
|\cM_{m;(i,k,l)}^{(0)}(\momt{(r)})|^2\Bigg]\,,
\label{eq:Sr10}
\\[3mm]
\cCS{ir_g}{r_g}{(1,0)}(\mom{}) \aand=
-(8\pi\as\mu^{2\eps})^2 c_{\Gamma}\frac{1}{s_{ir}}\frac{2\tzz{i}{r}}{\tzz{r}{i}}\bT_i^2
\label{eq:CirSr10}
\\&&\times
\Bigg[\left(\CA
\frac{1}{\eps^2} \frac{\pi\eps}{\sin({\pi\eps})}
\left(\frac{1}{s_{ir}}\frac{\tzz{i}{r}}{\tzz{r}{i}}\right)^\eps \cos(\pi\eps)
+ \frac{\beta_0}{2\eps}\frac{\mu^{-2\eps}S_\eps}{(4\pi)^2 c_\Gamma}\right)\Bigg]
|\cM_{m}^{(0)}(\momt{(r)})|^2\,.
\nn
\eeeq

The momentum fractions and eikonal functions were defined in
\eqns{eq:zt2}{eq:Sikr}, while the $m$ momenta
$\momt{(r)} \equiv \{\ti{p}_1,\ldots,\ti{p}_{m+1}\}$ ($p_r$ is absent)
entering the matrix elements on the right hand sides of
\eqnss{eq:Sr01}{eq:CirSr10} are defined as
\beq
\ti{p}_n^{\mu} =
\Lambda^{\mu}_{\nu}[Q,(Q-p_r)/\lambda_r] (p_n^{\nu}/\lambda_r)\,,
\qquad n\ne r\,,
\label{eq:PS_Sr}
\eeq
where
\beq
\lambda_r = \sqrt{1-y_{rQ}}\,,
\label{eq:lambdar}
\eeq
and
\beq
\Lambda^{\mu}_{\nu}[K,\wti{K}] = g^{\mu}_{\nu}
- \frac{2(K+\wti{K})^{\mu}(K+\wti{K})_{\nu}}{(K+\wti{K})^{2}} 
+ \frac{2K^{\mu}\wti{K}_{\nu}}{K^2}\,.
\label{eq:LambdaKKt}
\eeq
The matrix $\Lambda^{\mu}_{\nu}[K,\widetilde{K}]$ generates a (proper)
Lorentz transformation, provided $K^2 = \widetilde{K}^2 \ne 0$.
This momenum mapping leads to exact phase-space factorization
in the form of \eqn{eq:PS_fact}, where
\beq
[\rd p_{1;m}^{(r)}(p_r;Q)] =
\Jac{r}{1}(p_r;Q)
\,\frac{\rd^d p_r}{(2\pi)^{d-1}}\delta_{+}(p_r^2)\,,
\label{eq:dp_Sr}
\eeq
with Jacobian
\beq
\Jac{r}{1}(p_r;Q) = \lambda_{r}^{(m-1)(d-2)-2}\Theta(\lambda_{r})\,.
\eeq

Finally, we record the pole parts of the soft-type counterterms:
\beeq
\pole\cS{r_g}{(0,1)}(\mom{}) \aand= 
8\pi\as\mu^{2\eps}\sum_{i}\sum_{k\ne i} \frac14 \calS_{ik}(r)
\nn\\&&\times
\bra{m}{(0)}{(\momt{(r)})} 
\Big\{\bI(\momt{(r)};\eps), \bT_{i} \bT_{k}\Big\}
\ket{m}{(0)}{(\momt{(r)})}
\,,
\label{eq:polesSr01}
\\[3mm]
\pole\cCS{ir_g}{r_g}{(0,1)}(\mom{}) \aand= 
- 8\pi\as\mu^{2\eps} \frac{2}{s_{ir}}\frac{\tzz{i}{r}}{\tzz{r}{i}}\,\bT_i^2\,
\nn\\&&\times
\bra{m}{(0)}{(\momt{(r)})}\,\bI(\momt{(r)};\eps)\,\ket{m}{(0)}{(\momt{(r)})}
\,,
\label{eq:polesCirSr01}
\\[3mm]
\pole\cS{r_g}{(1,0)}(\mom{}) \aand=
8\pi\as\mu^{2\eps} \sum_{i}\sum_{k\ne i} \frac12 \calS_{ik}(r)
|\cM_{m;(i,k)}^{(0)}(\momt{(r)})|^2
\nn\\&&\times
\frac{\as}{2\pi} S_\eps\left[ \CA\left(
  \frac{1}{\eps^2} + \frac{1}{\eps} \ln\frac{y_{ik}}{y_{ir} y_{kr}}
+ \frac{1}{\eps} \ln\frac{\mu^2}{Q^2}
\right)
+ \frac{\beta_0}{2\eps}
\right]
\,,
\label{eq:polesSr10}
\\[3mm]
\pole\cCS{ir_g}{r_g}{(1,0)}(\mom{}) \aand=
-8\pi\as\mu^{2\eps} \frac{2}{s_{ir}}\frac{\tzz{i}{r}}{\tzz{r}{i}}\bT_i^2
|\cM_{m}^{(0)}(\momt{(r)})|^2
\nn \\&&\times
\frac{\as}{2\pi} S_\eps\left[ \CA\left(
  \frac{1}{\eps^2} + \frac{1}{\eps} \ln\frac{\tzz{i}{r}}{y_{ir} \tzz{r}{i}}
+ \frac{1}{\eps} \ln\frac{\mu^2}{Q^2}
\right)
+ \frac{\beta_0}{2\eps}
\right]
\,.
\label{eq:polesCirSr10}
\eeeq
The finite parts are given in \app{app:finiteparts}.

%
% Counterterms for the integrated approximate cross section
%

\section{Counterterms for the integrated approximate cross section}
\label{sec:dsigRRA1A1}

\subsection{Factorization in the collinear and soft limits}

We wish to construct the approximate cross section 
$\Big(\int_1\dsiga{RR}{1}_{m+2}\Big)\strut^{{\rm A}_{\scriptscriptstyle 1}}$
by the same procedure we used to construct $\dsiga{RV}{1}_{m+1}$,
therefore, we start by studying the infrared limits of 
$\bra{m+1}{(0)}{}\bIt(m,\eps)\ket{m+1}{(0)}{}$
when the momenta of a pair of partons becomes collinear or when a gluon
becomes soft.  

\subsubsection{Collinear limit}

In the limit when the momenta of partons $i$ and $r$ become collinear
(as defined precisely in \eqns{eq:sudakov}{eq:kTscaling}) we find%
\footnote{Note that the argument of the insertion operator on the right
hand side of \eqn{eq:CirMIM} is the same as the number of coloured
external legs in the matrix element.}
\beeq
&&
\bC{ir}\bra{m+1}{(0)}{(p_i,p_r,\ldots)}
\,\bIt(m,\eps)
\,\ket{m+1}{(0)}{(p_i,p_r,\ldots)} = 
8\pi\as\mu^{2\eps}
\nn\\&&\qquad
\times
\frac{1}{s_{ir}}
\Big(
  \bra{m}{(0)}{(p_{ir},\ldots)}
\,\bIt(m,\eps)\,\hP^{(0)}_{f_i f_r}
\,\ket{m}{(0)}{(p_{ir},\ldots)}
\nn\\&&\qquad\qquad
+ \R{ir}{(0)}(y_{ir},z_i y_{(ir)Q},z_r y_{(ir)Q};m,\eps)
\bra{m}{(0)}{(p_{ir},\ldots)}
\,\hP^{(0)}_{f_i f_r}\,
\ket{m}{(0)}{(p_{ir},\ldots)}\Big)
\,,
\label{eq:CirMIM}
\eeeq
where the function $\R{ir}{(0)}$ represents those terms that appear in
addition to the usual collinear factorization formula of the squared
matrix element, due to the presence of the insertion operator,
\beeq
&&
\R{ir}{(0)}(y_{ir},z_i y_{(ir)Q},z_r y_{(ir)Q};m,\eps) =
\aeps
\nn\\&&\qquad\times
\Bigg[
  \IcC{i}{(0)}(z_i y_{(ir)Q};m,\eps)\, \bT_i^2 
+ \IcC{r}{(0)}(z_r y_{(ir)Q};m,\eps)\, \bT_r^2
- \IcC{(ir)}{(0)}(y_{(ir)Q};m,\eps)\, \bT_{ir}^2
\nn\\&&\qquad\qquad
+ (\bT_{ir}^2 - \bT_i^2 - \bT_r^2)
\,\IcS{ir}{(0)}(y_{ir},z_i y_{(ir)Q},z_r y_{(ir)Q};m,\eps)
\Bigg]
\,.
\label{eq:Rir}
\eeeq
The $m$ parton matrix elements on the right hand side are obtained from
the $m+1$ parton matrix elements by removing partons $i$ and $r$ and
replacing them with a single parton denoted by $ir$ in the usual way.

Note that the existence of a universal collinear factorization formula as
given in \eqn{eq:CirMIM} is not guaranteed by the factorization
properties of QCD matrix elements, but depends also on the particular
definition of the subtraction term $\dsiga{RR}{1}_{m+2}$, which
determines the functional dependence of the insertion operator on the
momenta.  In being able to derive \eqn{eq:CirMIM} it is crucially
important that the part of $\bIt(m,\eps)$ that contains true
colour-correlations, that is $\IcS{il}{(0)}(y_{il},y_{iQ},y_{lQ};m,\eps)$,
depends on its arguments only through the combination
\beq
\frac{y_{il}}{y_{iQ} y_{lQ}}\,.
\eeq
This expression is independent of the momentum fractions in the
collinear limit $p_i || p_r$. Consequently, the functions
$\IcS{il}{(0)}$ and $\IcS{rl}{(0)}$ have the same limit as $p_i$ and
$p_r$ become collinear,
\beq
\bC{ir}\IcS{il}{(0)} = \bC{ir}\IcS{rl}{(0)}\,.
\label{eq:CirIcS}
\eeq
This is important because coherent soft-gluon emission from unresolved
partons implies that only the sum of
$\M{m+1;(i,l)}{(0)} + \M{m+1;(r,l)}{(0)}$ 
(or $\M{m+1;(j,i)}{(0)} + \M{m+1;(j,r)}{(0)}$) factorizes as
\beq
\bC{ir}
\left(\M{m+1;(i,l)}{(0)} + \M{m+1;(r,l)}{(0)} \right) =
8\pi\as\mu^{2\eps}\,\frac{1}{s_{ir}}
\,\bra{m}{(0)}{} \bT_{ir} \bT_l \,\hP^{(0)}_{f_i f_r} \ket{m}{(0)}{}\,.
\eeq
Then, if and only if \eqn{eq:CirIcS} is fulfilled, we can combine the
collinear limits as
\beq
\bC{ir}
\left(\IcS{il}{(0)}\M{m+1;(i,l)}{(0)} + \IcS{rl}{(0)}\M{m+1;(r,l)}{(0)}
\right) =
8\pi\as\mu^{2\eps}\,\IcS{(ir)l}{(0)}\,\frac{1}{s_{ir}}
\,\bra{m}{(0)}{} \bT_{ir} \bT_l \,\hP^{(0)}_{f_i f_r} \ket{m}{(0)}{}\,.
\eeq
The insertion operators used in other general NLO schemes do not possess
this property.

\subsubsection{Soft limit}

In computing the limit of $\bra{m+1}{(0)}{}\bIt(m,\eps)\ket{m+1}{(0)}{}$
as the momentum of parton $r$ becomes soft, we need the soft
factorization formula for the colour-correlated tree amplitudes as can
be found in \Ref{Somogyi:2005xz}. One finds
\beeq
&&
\bS{r}\bra{m+1}{(0)}{(p_r,\ldots)}
\,\bIt(m,\eps)
\,\ket{m+1}{(0)}{(p_r,\ldots)} = 
-8\pi\as\mu^{2\eps} \sum_{i} \sum_{k\ne i}
\frac12 \calS_{ik}(r)
\nn\\&&\quad\times
\Bigg(
\bra{m}{(0)}{(\ldots)}
\frac12 \{\bIt(m,\eps),\bT_i \bT_k\}
\ket{m}{(0)}{(\ldots)} 
\nn\\&&\quad\qquad
+ \R{ik,r}{(0)}(y_{ik},y_{ir},y_{kr},y_{iQ},y_{kQ},y_{rQ};m,\eps)
\SME{m;(i,k)}{0}{\ldots}\Bigg)
\label{eq:SrMIM}
\eeeq
if $r$ is a gluon and $\bS{r}\bra{m+1}{(0)}{}\bIt(m,\eps)\ket{m+1}{(0)}{}
= 0$ if $r$ is a quark or antiquark.  The $m$ parton matrix elements on
the right hand side are obtained from the original $m+1$ parton matrix
elements by simply removing parton $r$.
In \eqn{eq:SrMIM} the function
\beeq
&&
\R{ik,r}{(0)}(y_{ik},y_{ir},y_{kr},y_{iQ},y_{kQ},y_{rQ};m,\eps) =
\CA\, \aeps 
\Big(
\IcC{g}{(0)}(y_{rQ};m,\eps)
\nn\\&&\qquad
+ \IcS{ik}{(0)}(y_{ik},y_{iQ},y_{kQ};m,\eps)
- \IcS{ir}{(0)}(y_{ir},y_{iQ},y_{rQ};m,\eps)
- \IcS{rk}{(0)}(y_{rk},y_{rQ},y_{kQ};m,\eps)\Big)
\quad~
\label{eq:Rikr}
\eeeq
represents those terms that appear in addition to the usual soft
factorization formula of the squared matrix element due to the presence
of the insertion opeartor.

\subsubsection{Matching the collinear and soft limits}

The collinear limit of the soft factorization formula \eqn{eq:SrMIM} reads
\beeq
&&
\bC{ir}\bS{r}\bra{m+1}{(0)}{(p_i,p_r,\ldots)}
\,\bIt(m,\eps)
\,\ket{m+1}{(0)}{(p_i,p_r,\ldots)} =
8\pi\as\mu^{2\eps} \frac{2}{s_{ir}}\frac{z_i}{z_r}
\,\bT_i^2
\nn\\&&\qquad\qquad\times
\Big(\bra{m}{(0)}{(p_i,\ldots)}\,\bIt(m,\eps)\,\ket{m}{(0)}{(p_i,\ldots)}
+ \SME{m}{0}{p_i,\ldots}
\bC{ir} \R{ik,r}{(0)}
\Big)\,,
\label{eq:CirSrMIM}
\eeeq
with
\beeq
&&
\bC{ir} \R{ik,r}{(0)}(y_{ik},y_{ir},y_{kr},y_{iQ},y_{kQ},y_{rQ};m,\eps) =
\CA\aeps
\nn\\&&\qquad\qquad\times
\Big(
\IcC{g}{(0)}(z_r y_{(ir)Q};m,\eps) 
- \IcS{ir}{(0)}(y_{ir},z_i y_{(ir)Q},z_r y_{(ir)Q};m,\eps)\Big)\,.
\label{eq:CirRikr}
\eeeq
The soft limit of the collinear factorization formula \eqn{eq:CirMIM} is
\beeq
&&
\bS{r}\bC{ir}\bra{m+1}{(0)}{(p_i,p_r,\ldots)}
\,\bIt(m,\eps)
\,\ket{m+1}{(0)}{(p_i,p_r,\ldots)} =
8\pi\as\mu^{2\eps} \frac{2}{s_{ir}}\frac{1}{z_r}
\,\bT_i^2
\nn\\&&\qquad\qquad\times
\Big(\bra{m}{(0)}{(p_i,\ldots)}\,\bIt(m,\eps)\,\ket{m}{(0)}{(p_i,\ldots)}
+ \SME{m}{0}{p_i,\ldots}
\bS{r} \R{ir}{(0)}
\Big)\,.
\label{eq:SrCirMIM}
\eeeq
with
\beeq
&&
\bS{r} \R{ir}{(0)}(y_{ir},z_i y_{(ir)Q},z_r y_{(ir)Q};m,\eps) =
\CA\aeps
\nn\\&&\qquad\qquad\times
\Big(
\IcC{g}{(0)}(z_r y_{(ir)Q};m,\eps) 
- \IcS{ir}{(0)}(y_{ir},y_{(ir)Q},z_r y_{(ir)Q};m,\eps)\Big)\,.
\label{eq:SrRir}
\eeeq
Thus the same arguments as below \eqn{eq:SrCirRVnew} apply here as well,
therefore,
\beeq
&&
\bC{ir}(\bS{r}-\bC{ir}\bS{r})\bra{m+1}{(0)}{}\bIt(m,\eps)\ket{m+1}{(0)}{}=0\,,
\\ &&
\bS{r}(\bC{ir}-\bC{ir}\bS{r})\bra{m+1}{(0)}{}\bIt(m,\eps)\ket{m+1}{(0)}{}=0
\eeeq
and our candidate counterterm has the same structure as in \eqn{eq:RV_A1},
\beeq
&&
\bA{1}\,\bra{m+1}{(0)}{}\bIt(m,\eps)\ket{m+1}{(0)}{} = 
\nn\\ &&\quad
= \sum_r \left[\sum_{i\ne r} \frac12 \bC{ir}
+\left(\bS{r}-\sum_{i\ne r} \bCS{ir}{r}\right)\right]
\left(\SME{m+1}{0}{p_i,p_r,\ldots}\otimes \bIt(m,\eps)\right)
\,.
\quad~
\label{eq:MIM_A1}
\eeeq
As before the cancellation of the collinear terms in the soft limit
requires the symmetry factor multiplying the collinear term, but
not the collinear-soft one.

\subsection{Counterterms}
\label{sec:A1A1counterterms}

In order to define the counterterms, we extend \eqn{eq:MIM_A1} over the
whole phase space as done in \sect{sec:RVcounterterms}. We introduce the
phase space factorization as in \eqn{eq:PS_fact} and write the subtraction
term that regularizes the kinematical singularities of the integrated
approximate cross section as
\beq
\Big(\int_1\dsiga{RR}{1}_{m+2}\Big)\strut^{{\rm A}_{\scriptscriptstyle 1}} =
\PS{m}{}[\rd p_1] \bcA{1}{(0\otimes I)}
\Big(\M{m+1}{(0)}\otimes\bIt(m,\eps) \Big)
\,,
\label{eq:dsigRRA11}
\eeq
where
\beeq
&&
\bcA{1}{(0\otimes I)}\Big(\M{m+1}{(0)}\otimes\bIt(m,\eps) \Big) =
\nn\\&&\qquad
= \sum_{r} \left[\sum_{i\ne r} \frac{1}{2} \cC{ir}{(0,0 \otimes I)}(\mom{})
+ \left(\cS{r}{(0,0 \otimes I)}(\mom{})
- \sum_{i\ne r} \cCS{ir}{r}{(0,0 \otimes I)}(\mom{})\right) 
\right]
\nn\\&&\qquad
+ \sum_{r} \left[\sum_{i\ne r} \frac{1}{2} \cC{ir}{R\times(0,0)}(\mom{})
+ \left(\cS{r}{R\times(0,0)}(\mom{}) 
- \sum_{i\ne r} \cCS{ir}{r}{R\times(0,0)}(\mom{})\right) 
\right]\,.
\label{eq:A10M0I}
\eeeq
We now define all terms on the right hand side of this equation
precisely.  The structure of \eqn{eq:A10M0I} is the same as that of
\eqn{eq:A11M0M1}.  Thus defining true subtraction terms starting from
the limiting formulae of the previous subsection follows the steps of
\sect{sec:RVcounterterms}.

\subsubsection{Collinear counterterms}

The collinear subtraction terms read
\beeq
&&
\cC{ir}{(0,0 \otimes I)}(\mom{}) =
8\pi\as\mu^{2\eps}\frac{1}{s_{ir}}
\nn\\&&\qquad\times
\bra{m}{(0)}{(\momt{(ir)})}\,\bIt(\momt{(ir)};m,\eps)
\,\hP^{(0)}_{f_i f_r}(\tzz{i}{r},\tzz{r}{i},\kTt{i,r};\eps)
\,\ket{m}{(0)}{(\momt{(ir)})}
\label{eq:Cir00xI}
\eeeq
and
\beeq
\cC{ir}{R\times(0,0)}(\mom{}) \aand=
8\pi\as\mu^{2\eps}\,\frac{1}{s_{ir}}
\,\R{ir}{(0)}(y_{ir},\tzz{i}{r}y_{\wti{ir}Q},\tzz{r}{i}y_{\wti{ir}Q};m,\eps)
\nn\\&&\times
\bra{m}{(0)}{(\momt{(ir)})}
\hP^{(0)}_{f_i f_r}(\tzz{i}{r},\tzz{r}{i},\kTt{i,r};\eps)
\ket{m}{(0)}{(\momt{(ir)})}
\,.
\label{eq:CirRx00}
\eeeq
Computing the pole parts of \eqns{eq:Cir00xI}{eq:CirRx00}, we can easily
see that
\beq
\pole\Big[\cC{ir}{(0,0 \otimes I)}(\mom{}) + \cC{ir}{(0,1)} (\mom{})\Big] = 0
\label{eq:polesCir00xI}
\eeq
and
\beq
\pole\Big[\cC{ir}{R\times(0,0)}(\mom{}) + \cC{ir}{(1,0)} (\mom{})\Big] = 0
\,.
\label{eq:polesCirRx00}
\eeq
Therefore, the sum of \eqns{eq:Cir01}{eq:Cir00xI}
as well as that of \eqns{eq:Cir10}{eq:CirRx00} is finite in
$d = 4$ dimensions.

\subsubsection{Soft-type counterterms}

The soft-type counterterms are defined as
\beeq
\cS{r_g}{(0,0\otimes I)}(\mom{}) \aand=
-8\pi\as\mu^{2\eps}\sum_{i}\sum_{k\ne i}\frac14 \calS_{ik}(r)
\nn\\ &&\times
\,\bra{m}{(0)}{(\momt{(r)})}
\{\bIt(\momt{(r)};m,\eps),\bT_i \bT_k\}
\ket{m}{(0)}{(\momt{(r)})}\,,
\label{eq:Sr00xI}
\\[3mm]
\cC{ir_g}{}\cS{r_g}{(0,0\otimes I)}(\mom{}) \aand=
8\pi\as\mu^{2\eps} \frac{1}{s_{ir}}\frac{2\tzz{i}{r}}{\tzz{r}{i}}
\nn\\ &&\times
\bT_i^2\,\bra{m}{(0)}{(\momt{(r)})}
\bIt(\momt{(r)};m,\eps)\ket{m}{(0)}{(\momt{(r)})}\,,
\label{eq:CirSr00xI}
\\[3mm]
\cS{r_g}{R\times(0,0)}(\mom{}) \aand=
-8\pi\as\mu^{2\eps}\sum_{i}\sum_{k\ne i}
\frac12 \calS_{ik}(r)\,\SME{m;(i,k)}{0}{\momt{(r)}}
\nn\\ &&\times
\,\R{ik,r}{(0)}(y_{ik},y_{ir},y_{kr},y_{iQ},y_{kQ},y_{rQ};m,\eps)
\label{eq:SrRx00}
\\[3mm]
\cC{ir_g}{}\cS{r_g}{R\times(0,0)}(\mom{}) \aand=
8\pi\as\mu^{2\eps}\frac{1}{s_{ir}}\frac{2\tzz{i}{r}}{\tzz{r}{i}}
\bT_i^2\SME{m}{0}{\momt{(r)}}\CA\aeps
\nn\\&&\times
\,\Big(
  \IcC{g}{(0)}(\tzz{r}{i} y_{(ir)Q};m,\eps) 
- \IcS{ir}{(0)}(y_{ir},\tzz{i}{r} y_{(ir)Q},\tzz{r}{i} y_{(ir)Q};m,\eps)\Big)
\,.
\label{eq:CirSrRx00}
\eeeq
We can simplify the arguments in the second line of \eqn{eq:CirSrRx00}
using \eqn{eq:zt2},
\beeq
&&
  \IcC{g}{(0)}(\tzz{r}{i} y_{(ir)Q};m,\eps) 
- \IcS{ir}{(0)}(y_{ir},\tzz{i}{r} y_{(ir)Q},\tzz{r}{i} y_{(ir)Q};m,\eps)
\nn \\ &&\qquad\qquad
= \IcC{g}{(0)}(y_{rQ};m,\eps) 
- \IcS{ir}{(0)}(y_{ir},y_{iQ}, y_{rQ};m,\eps)\,.
\eeeq
Similarly to the collinear cases, the pole parts cancel term by term
between \eqnss{eq:Sr01}{eq:CirSr10}
and \eqnss{eq:Sr00xI}{eq:CirSrRx00},
\beeq
&&
\pole\Big[\cS{r}{(0,0 \otimes I)}(\mom{}) +
\cS{r}{(0,1)} (\mom{})\Big] = 0
\,,
\label{eq:polesSr00xI}
\\[3mm]
&&
\pole\Big[\cC{ir}{}\cS{r}{(0,0\otimes I)}(\mom{}) +
\cC{ir}{}\cS{r}{(0,1)} (\mom{})\Big] = 0
\,,
\label{eq:polesCirSr00xI}
\\[3mm]
&&
\pole\Big[\cS{r}{R\times(0,0)}(\mom{}) +
\cS{r}{(1,0)} (\mom{})\Big] = 0
\,,
\label{eq:polesSrRx00}
\\[3mm]
&&
\pole\Big[\cC{ir}{}\cS{r}{R\times(0,0)}(\mom{}) +
\cC{ir}{}\cS{r}{(1,0)} (\mom{})\Big] = 0
\,.
\label{eq:polesCirSrRx00}
\eeeq
Consequently, the sum of \eqns{eq:Sr01}{eq:Sr00xI},
that of \eqns{eq:CirSr01}{eq:CirSr00xI},
that of \eqns{eq:Sr10}{eq:SrRx00} and that of
\eqns{eq:CirSr10}{eq:CirSrRx00} are finite in $d = 4$ dimensions.

We help the reader in grasping the various cancellations by visualizing
the subtraction terms in \fig{fig:RV-RVA+A1-A11}.  The first picture
corresponds to the squared matrix element of the $m+1$ final-state
partons. The fully shaded circle represents the Born amplitude, while
the circle with a whole is the one-loop amplitude. The following terms
in the first squared brackets correspond to the terms that build
$\bcA{1}{(1)} \,2\Real \bra{m+1}{(0)}{}\ket{m+1}{(1)}{}$, defined in
\eqn{eq:A11M0M1}; the first one represents (0,1)-type terms like
$\cC{ir}{(0,1)}(\mom{})$ and the second is for (1,0)-type terms such as
$\cC{ir}{(1,0)}(\mom{})$.  The first picture in the second line is the
result of the integration in \eqn{eq:I1dsigRRA1_fin2}. Finally, the
terms in the second squared brackets represent terms that contribute
to $\bcA{1}{(0\otimes I)}\Big(\M{m+1}{(0)}\otimes\bIt(m,\eps) \Big)$,
defined in \eqn{eq:A10M0I}; the first representing terms of the
$(0,0\otimes \bI)$-type, such as $\cC{ir}{(0,0 \otimes I)}(\mom{})$,
while the second one stands for $R\times(0,0)$-type terms like
$\cC{ir}{R\times(0,0)}(\mom{})$.  The factorized one-parton factors
correspond to Altarelli-Parisi kernels, with azimuthal correlations
included, or eikonal factors, with colour-correlations included, either
at tree-level (fully shaded circles), or at one-loop (circles with
holes), or $R$-functions (boxes).
\definecolor{mygrey}{rgb}{0.64,0.66,0.62}
\definecolor{mygreyd}{rgb}{0.40,0.40,0.40}
\begin{figure}
\begin{center}
\begin{pspicture}(0,0)(16,4)

\scalebox{0.45}{%

\SpecialCoor
\psarc(3.5,3){2.69}{(2.5,1)}{(-2.5,1)}
\psarc(3.5,5){2.69}{(-2.5,-1)}{(2.5,-1)}
\psline(1,4)(6,4)
\psline[linestyle=dashed](3.5,2)(3.5,6)
\psdots*[dotscale=0.3](3.3,5.2)(3.3,4.8)(3.3,4.4)
\psdots*[dotscale=0.3](3.3,3.6)(3.3,3.2)(3.3,2.8)
\pscircle[fillstyle=solid,fillcolor=mygrey](1,4){0.9}
\pscircle[fillstyle=solid,fillcolor=white](1,4){0.5}\rput(1,4){\LARGE $m\!+\!1$}
\pscircle[fillstyle=solid,fillcolor=mygrey](6,4){0.9}\rput(6,4){\LARGE $m\!+\!1$}
\rput[l](4,4.3){\LARGE $r$}

% A1

\rput(8,4){\Huge $- \,\,\Bigg[$}

\psarc[origin={-9,0}](3.5,3){2.69}{(2.5,1)}{(-2.5,1)}
\psarc[origin={-9,0}](3.5,5){2.69}{(-2.5,-1)}{(2.5,-1)}
\psline[origin={-9,0},linestyle=dashed](3.5,2)(3.5,6)
\psdots*[origin={-9,0},dotscale=0.3](3.3,4.6)(3.3,4)(3.3,3.4)
\pscircle[origin={-9,0},fillstyle=solid,fillcolor=mygrey](1,4){0.9}
\pscircle[origin={-9,0},fillstyle=solid,fillcolor=white](1,4){0.5}\rput(10,4){\LARGE $m$}
\pscircle[origin={-9,0},fillstyle=solid,fillcolor=mygrey](6,4){0.9}\rput(15,4){\LARGE $m$}

\rput(16.7,4){\Huge $\otimes$}
\pscircle[fillstyle=solid,fillcolor=mygrey](18,4){0.5}
\psline(18.5,4)(19.5,4)\uput[u](19,4){\LARGE $r$}

\rput(20.2,4){\Huge $\!+\!$}

\psarc[origin={-21,0}](3.5,3){2.69}{(2.5,1)}{(-2.5,1)}
\psarc[origin={-21,0}](3.5,5){2.69}{(-2.5,-1)}{(2.5,-1)}
\psline[origin={-21,0},linestyle=dashed](3.5,2)(3.5,6)
\psdots*[origin={-21,0},dotscale=0.3](3.3,4.6)(3.3,4)(3.3,3.4)
\pscircle[origin={-21,0},fillstyle=solid,fillcolor=mygrey](1,4){0.9}\rput(22,4){\LARGE $m$}
\pscircle[origin={-21,0},fillstyle=solid,fillcolor=mygrey](6,4){0.9}\rput(27,4){\LARGE $m$}

\rput(28.7,4){\Huge $\otimes$}
\pscircle[fillstyle=solid,fillcolor=mygrey](30,4){0.5}
\pscircle[fillstyle=solid,fillcolor=white](30,4){0.3}
\psline(30.5,4)(31.5,4)\uput[u](31,4){\LARGE $r$}

\rput[l](32,4){\Huge $\!+\! \ldots\,\,\Bigg]$}

% <M|I|M>

\rput(-0.5,-1){\Huge $\!+\!$}

\SpecialCoor
\psarc[origin={0,5}](3.5,3){2.69}{(2.5,1)}{(-2.5,1)}
\psarc[origin={0,5}](3.5,5){2.69}{(-2.5,-1)}{(2.5,-1)}
\psline[origin={0,5}](1,4)(6,4)
\psframe[origin={0,5},fillstyle=solid,fillcolor=mygrey](3,2)(4,6)
\rput(3.5,-1){\LARGE $\mbox{\boldmath{$I$}}_{m}$}
\psdots*[origin={0,5},dotscale=0.3](2.8,5.2)(2.8,4.8)(2.8,4.4)
\psdots*[origin={0,5},dotscale=0.3](2.8,3.6)(2.8,3.2)(2.8,2.8)
\pscircle[origin={0,5},fillstyle=solid,fillcolor=mygrey](1,4){0.9}\rput(1,-1){\LARGE $m\!+\!1$}
\pscircle[origin={0,5},fillstyle=solid,fillcolor=mygrey](6,4){0.9}\rput(6,-1){\LARGE $m\!+\!1$}
\rput[l](4.5,-0.7){\LARGE $r$}

% IA1

\rput(8,-1){\Huge $- \,\,\Bigg[$}

\psarc[origin={-9,5}](3.5,3){2.69}{(2.5,1)}{(-2.5,1)}
\psarc[origin={-9,5}](3.5,5){2.69}{(-2.5,-1)}{(2.5,-1)}
\psframe[origin={-9,5},fillstyle=solid,fillcolor=mygrey](3,2)(4,6)
\rput(12.5,-1){\LARGE $\mbox{\boldmath{$I$}}_{m}$}
\psdots*[origin={-9,5},dotscale=0.3](2.8,4.6)(2.8,4)(2.8,3.4)
\pscircle[origin={-9,5},fillstyle=solid,fillcolor=mygrey](1,4){0.9}\rput(10,-1){\LARGE $m$}
\pscircle[origin={-9,5},fillstyle=solid,fillcolor=mygrey](6,4){0.9}\rput(15,-1){\LARGE $m$}

\rput(16.7,-1){\Huge $\otimes$}
\pscircle[fillstyle=solid,fillcolor=mygrey](18,-1){0.5}
\psline(18.5,-1)(19.5,-1)\uput[u](19,-1){\LARGE $r$}

\rput(20.2,-1){\Huge $\!+\!$}

\psarc[origin={-21,5}](3.5,3){2.69}{(2.5,1)}{(-2.5,1)}
\psarc[origin={-21,5}](3.5,5){2.69}{(-2.5,-1)}{(2.5,-1)}
\psline[origin={-21,5},linestyle=dashed](3.5,2)(3.5,6)
\psdots*[origin={-21,5},dotscale=0.3](3.3,4.6)(3.3,4)(3.3,3.4)
\pscircle[origin={-21,5},fillstyle=solid,fillcolor=mygrey](1,4){0.9}\rput(22,-1){\LARGE $m$}
\pscircle[origin={-21,5},fillstyle=solid,fillcolor=mygrey](6,4){0.9}\rput(27,-1){\LARGE $m$}

\rput(28.7,-1){\Huge $\otimes$}
\psframe[fillstyle=solid,fillcolor=mygrey](29.5,-1.5)(30.5,-0.5)
\rput(30,-1){\LARGE $\mathrm{R}$}
\psline(30.5,-1)(31.5,-1)\uput[u](31,-1){\LARGE $r$}

\rput[l](32,-1){\Huge $\!+\! \ldots\,\,\Bigg]$}
}
\end{pspicture} \end{center}
~\vskip 6mm
\caption{Graphical representation of the squared matrix element and the
subtraction terms.}
\label{fig:RV-RVA+A1-A11}
\end{figure}
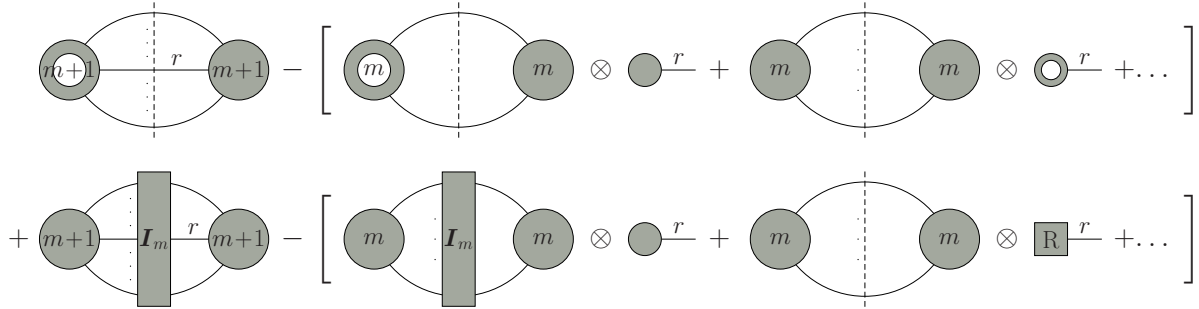

The cancellations of the $\eps$ poles occurs vertically, term by term as
shown previously. The regularization of the kinematical singularities
takes place horizontally, between the first term and the terms in the
following brackets, separately in each line. Kinematical singularities,
introduced by the subtractions terms, are screened by the jet function,
$J_m$, just as in NLO computations.

We conclude that the $(m+1)$-parton contribution in \eqn{eq:sigmaNNLOm+1}
is free of $\eps$ poles as well as unscreened kinematical singularities.
We can set $\eps = 0$ and compute the integral of $\dsig{NNLO}_{m+1}$
by standard Monte Carlo methods. Such an integration uses the finite
parts in the $\eps$-expansion of the subtraction terms as $\eps \to 0$.
The finite parts of the integrated approximate cross section of
\eqn{eq:I1dsigRRA1_fin2} can be found in \Ref{Somogyi:2006_1},%
\footnote{Notice the shift of $m$ by one unit in this paper ($m \to m+1$)
as compared to the value in \Ref{Somogyi:2006_1}.}
while those of \eqn{eq:A11M0M1} together with the finite parts of the
counterterms in \eqn{eq:A10M0I} are presented in \app{app:finiteparts}.

\section{Numerical results}
\label{sec:results}

In this paper we have defined explicitly all subtraction terms that are
needed to make $\dsig{RV}_{m+1}$ integrable in $d = 4$ dimensions for
processes without coloured partons in the initial state. We have
proven the cancellation of the $\eps$ poles explicitly. In order to 
demonstrate that the subtraction terms indeed regularize the cross
section for the real-virtual correction, we consider the contribution to
the theoretical predictions for the three-jet event-shape distributions
thrust ($T$) and $C$-parameter in electron-positron annihilation, when
the jet function is a functional
\beq
J_n(p_1,\ldots,p_n;O) = \delta(O-O_3(p_1,\ldots,p_n))\:,
\eeq
with $O_3(p_1,\ldots,p_n)$ being the value of either $\tau \equiv 1-T$
or $C$ for a given event $(p_1,\ldots,p_n)$.  
Starting from randomly chosen phase space points and approaching the
singly-unresolved (soft and/or collinear) regions of the phase space in
successive steps, we have checked numerically that the sum of the
subtraction terms has the same limits (up to integrable square-root
singularities) as the squared matrix element itself.

The perturbative expansion of the $n^{\rm th}$ moment of a three-jet
observable at a fixed scale $\mu = Q$ and NNLO accuracy can be
parametrised as 
\beeq
\la O^n \ra \aand\equiv
\int\!\rd O\,O^n\,\frac{1}{\sigma_0}\frac{\dsig{}}{\rd O}(O) =
\nn\\[3mm]&&=
  \left(\frac{\as(Q)}{2\pi}\right)   A_{O}^{(n)}
+ \left(\frac{\as(Q)}{2\pi}\right)^2 B_{O}^{(n)}
+ \left(\frac{\as(Q)}{2\pi}\right)^3 C_{O}^{(n)}\:,
\eeeq
where according to \eqn{eq:sigmaNNLOfin}, the NNLO correction is a sum
of three contributions,
\beq
C_{O}^{(n)} =
C_{O;5}^{(n)} + C_{O;4}^{(n)} + C_{O;3}^{(n)}\:.
\eeq
Carrying out the phase space integrations in \eqn{eq:sigmaNNLOm+1}, we
computed the four-parton contribution $C_{O;4}^{(n)}$ as defined in
this article. The predictions for the first three moments of $\tau$ and
the $C$-parameter, obtained using about four million Monte Carlo
events, are presented in \tab{tab:C}.  In performing the numerical
integrations, we do not encounter more severe numerical problems than
known from computing the real-emission contribution in NLO computations
and the computation of differential distributions does not pose any
problem. The required CPU time is however much longer because of the
much more cumbersome expressions that enter the various loop matrix elements.
\begin{table}
\begin{center}
\caption{The moments $C_{\tau;4}^{(n)}$ and $C_{C;4}^{(n)}$.}
\label{tab:C}
\begin{tabular}{|c|c|c|}
\hline
\hline
 & & \\[-4mm]
 $n$ & $C_{\tau;4}^{(n)}$ & $C_{C;4}^{(n)}$ \\
 & & \\[-4mm]
\hline
 & & \\[-4mm]
  1  & -\res{1.23}{0.01}{ 3} & -\res{4.33}{0.05}{ 3} \\
  2  & -\res{2.55}{0.01}{ 2} & -\res{3.25}{0.02}{ 3} \\
  3  & -\res{4.79}{0.03}{ 1} & -\res{1.80}{0.01}{ 3} \\
\hline
\hline
\end{tabular}
\end{center}
\end{table}

\section{Conclusions}
\label{sec:conclusions}

In a companion paper \cite{Somogyi:2006_2} we set up a subtraction
scheme for computing NNLO corrections to QCD jet cross sections to
processes without coloured partons in the initial state. The scheme is
completely general in the sense that any number of massless coloured
final-state partons (massive vector bosons are assumed to decay into
massless fermions) are allowed provided the necessary squared matrix
elements are known. 

Three types of corrections contribute to the NNLO corrections: the
doubly-real, the real-virtual and the doubly-virtual ones. In
\Ref{Somogyi:2006_2} we defined the subtraction terms for the doubly-real
emissions; those to the real-virtual correction can be found in the
present paper. By rendering these two contributions finite in $d = 4$
dimensions, the KLN theorem ensures that for infrared safe observables
adding the subtractions above to the doubly-virtual correction that
becomes also finite in $d = 4$ dimensions.

The subtraction terms for the real-virtual corrections presented here
are local in $d = 4 - 2 \eps$ dimensions, include complete colour and
azimuthal correlations. The expressions were derived by extending the
singly-unresolved limits of the one-loop squared matrix elements over
the whole phase space and also extending the singly-unresolved limits
of the integrated approximate cross section, used for regularizing the
kinematical singularities of the cross section for doubly-real
emmissions over the singly-unresolved regions of the phase space.

In order to demonstrate that the subtracted cross section is indeed
integrable, we have computed the corresponding contributions to the
first three moments of two three-jet event-shape observables, the
thrust and the $C$-parameter. In performing the numerical integrations,
we do not encounter more severe numerical problems than known from NLO
computations. The required CPU time is however much longer because of
the much more cumbersome expressions that enter the various loop matrix
elements.  

\section*{Acknowledgments}
We are grateful to Vittorio Del Duca for his comments on the manuscript
and for the hospitality of the CERN Theory Division, where this work
was completed.
This research was supported in part by the Hungarian Scientific Research
Fund grant OTKA K-60432.

\appendix

\section{Finite parts of the subtraction terms}
\label{app:finiteparts}

In this appendix we present the finite parts, as defined in the
decomposition (\ref{eq:pole+fin}), of \eqns{eq:A11M0M1}{eq:A10M0I}, term
by term. We start with the collinear counterterms.
The finite part of the sum of \eqns{eq:Cir01}{eq:Cir00xI} is
\beeq
\finite
\aand\!\!
\Big[\cC{ir}{(0,1)}(\mom{}) + \cC{ir}{(0,0\otimes I)}(\mom{})\Big] =
4\as^2\,\frac{1}{s_{ir}}
\nn\\&& \times
\Bigg[
\,2\Real \bra{m}{(0)}{(\momt{(ir)})}
\,\hP_{f_i f_r}^{(0)}(\tzz{i}{r},\tzz{r}{i},\kTt{i,r})
\,\finite\ket{m}{(1)}{(\momt{(ir)})}
\nn\\&& \qquad
+ \sum_i \Bigg( \sum_{k\ne i}
\finite\Big[
  \IcS{ik}{(0)}(y_{\ti{i}\ti{k}},y_{\ti{i}Q},y_{\ti{k}Q};m)
+ \IcCS{(0)}(m)
\Big]
\nn\\&& \qquad\qquad\qquad\times
\bra{m}{(0)}{(\momt{(ir)})}
\,\bT_i \bT_k 
\,\hP_{f_i f_r}^{(0)}(\tzz{i}{r},\tzz{r}{i},\kTt{i,r})
\,\ket{m}{(0)}{(\momt{(ir)})}
\nn\\&& \qquad\qquad\quad
+ \finite\Big[
  \IcC{i}{(0)}(y_{\ti{i}Q};m)
+ \IcCS{(0)}(m)
\Big]
\nn\\&& \qquad\qquad\qquad\times
\,\bT_i^2
\bra{m}{(0)}{(\momt{(ir)})}
\,\hP_{f_i f_r}^{(0)}(\tzz{i}{r},\tzz{r}{i},\kTt{i,r})
\,\ket{m}{(0)}{(\momt{(ir)})}
\Bigg)
\Bigg]
\,,
\label{eq:finCir01+finCir00xI}
\eeeq
where the finite part of the one-loop amplitude is defined by the
following equation:
\beq
\ket{m}{(1)}{(\mom{})} =
- \frac12 \bI(\mom{};\eps) \ket{m}{(0)}{(\mom{})}
+ \frac{\as}{2\pi} \finite \ket{m}{(1)}{(\mom{})}
\,,
\eeq
with $\bI(\mom{};\eps)$ given in \eqn{eq:bI}.  The finite parts of the
functions $\IcS{ik}{(0)} + \IcCS{(0)}$ and $\IcC{i}{(0)}+\IcCS{(0)}$
can be found in the appendix of \Ref{Somogyi:2006_1}.
Next we turn to the finite parts of \eqns{eq:Cir10}{eq:CirRx00}, which
are given separately for the various flavour combinations.
For $f_i=f_r=g$:
\beeq
\finite
\aand\!\!
\Big[\cC{g_ig_r}{(1,0)}(\mom{}) + \cC{g_ig_r}{R\times(0,0)}(\mom{})\Big] =
\CA\,4 \as^2\,\frac{1}{s_{ir}}
\nn\\&& \times
\Bigg[
\Bigg(
\finite\Big[
  \frac12
\left(\IcC{gg}{(0)}(\tzz{i}{r} y_{\wti{ir}Q}; m)
    + \IcC{gg}{(0)}(\tzz{r}{i} y_{\wti{ir}Q}; m)
    - \IcC{gg}{(0)}(y_{\wti{ir}Q}; m)
\right)
\nn\\[1mm] && \qquad\qquad
+ \Nf
\left(\IcC{q\qb}{(0)}(\tzz{i}{r} y_{\wti{ir}Q}; m)
    + \IcC{q\qb}{(0)}(\tzz{r}{i} y_{\wti{ir}Q}; m)
    - \IcC{q\qb}{(0)}(y_{\wti{ir}Q}; m)
\right)
\nn\\[1mm] && \qquad\qquad
- \IcS{ir}{(0)}(y_{ir},\tzz{i}{r} y_{\wti{ir}Q},\tzz{r}{i} y_{\wti{ir}Q}; m)
- \IcCS{(0)}(m)
\Big]
\nn\\[1mm] && \qquad\qquad
- \frac12 \ln^2\frac{\tzz{i}{r}}{\tzz{r}{i}}
- \frac12 \ln^2 y_{ir}
- \ln y_{ir} \ln(\tzz{i}{r} \tzz{r}{i})
+ \frac{\pi^2}{3}
+ \frac{\beta_0}{2\CA}\ln\frac{\mu^2}{Q^2}
\Bigg)
\nn\\[1mm] && \qquad \times
\bra{m}{(0)}{(\momt{(ir)})}
\,\hP_{g_i g_r}^{(0)}(\tzz{i}{r},\tzz{r}{i},\kTt{i,r})
\,\ket{m}{(0)}{(\momt{(ir)})}
\nn\\[1mm] && \qquad
- \frac23 (\CA - 2 \TR\Nf)
\,\bra{m}{(0)}{(\momt{(ir)})}
\,\frac{\kTt{i,r}^\mu\kTt{i,r}^\nu}{\kTt{i,r}^2}
\,\ket{m}{(0)}{(\momt{(ir)})}
\Bigg]
\,.
\qquad~
\label{eq:finCgg10+finCggRx00}
\eeeq
For $f_i=q$, $f_r=\qb$:
\beeq
\finite
\aand\!\!
\Big[\cC{q_i\qb_r}{(1,0)}(\mom{}) + \cC{q_i\qb_r}{R\times(0,0)}(\mom{})
\Big] =
4 \as^2\,\frac{1}{s_{ir}}
\nn\\&& \times
\Bigg(
\CF\,\finite\Big[
  \IcC{qg}{(0)}(\tzz{i}{r} y_{\wti{ir}Q}; m)
+ \IcC{qg}{(0)}(\tzz{r}{i} y_{\wti{ir}Q}; m)
\Big]
\nn\\[1mm] && \qquad
- \CA\,\finite\Big[
  \frac12 \IcC{gg}{(0)}(y_{\wti{ir}Q}; m)
+ \Nf \IcC{q\qb}{(0)}(y_{\wti{ir}Q}; m)
\Big]
\nn\\[1mm] && \qquad
+ (\CA-2\CF)\,\finite\Big[
  \IcS{ir}{(0)}(y_{ir},\tzz{i}{r} y_{\wti{ir}Q},\tzz{r}{i} y_{\wti{ir}Q}; m)
+ \IcCS{(0)}(m)
\Big]
\nn\\[1mm] && \qquad
+ (\CA - 2\CF) \left(\frac12 \ln^2 y_{ir} - \frac{\pi^2}{2} + 1\right)
+ 2 (\gamma_g - \gamma_q) (2 - \ln y_{ir})
\nn\\[1mm] && \qquad
- \CA\left(
  \frac12 \ln^2\frac{\tzz{i}{r}}{\tzz{r}{i}}
+ \ln y_{ir}\,\ln(\tzz{i}{r}\,\tzz{r}{i}) + \frac{\pi^2}{6}
\right)
+ \frac19 (\CA + 4\TR\Nf)
\nn\\[1mm] &&  \qquad
+ \frac{\beta_0}{2}\ln\frac{\mu^2}{Q^2}
\Bigg)
\bra{m}{(0)}{(\momt{(ir)})}
\,\hP_{q_i \qb_r}^{(0)}(\tzz{i}{r},\tzz{r}{i},\kTt{i,r})
\,\ket{m}{(0)}{(\momt{(ir)})}
\,.
\label{eq:finCqq10+finCqqRx00}
\eeeq
For $f_i=q$, $f_r=g$:
\beeq
\finite
\aand\!\!
\Big[\cC{q_ig_r}{(1,0)}(\mom{}) + \cC{q_ig_r}{R\times(0,0)}(\mom{}) \Big] =
4 \as^2 \frac{1}{s_{ir}}
\nn\\ && \times
\Bigg[
\Bigg(
\CF\,\finite\Big[
  \IcC{qg}{(0)}(\tzz{i}{r} y_{\wti{ir}Q}; m)
- \IcC{qg}{(0)}(y_{\wti{ir}Q}; m)
\Big]
\nn\\[1mm] && \qquad
+ \CA\,\finite\Big[
  \frac12 \IcC{gg}{(0)}(\tzz{r}{i} y_{\wti{ir}Q}; m)
+ \Nf \IcC{q\qb}{(0)}(\tzz{r}{i} y_{\wti{ir}Q}; m)
\Big]
\nn\\[1mm] && \qquad
- \CA\,\finite\Big[
  \IcS{ir}{(0)}(y_{ir},\tzz{i}{r} y_{\wti{ir}Q},\tzz{r}{i} y_{\wti{ir}Q}; m)
+ \IcCS{(0)}(m)
\Big]
\nn\\[1mm] && \qquad
+ 2\CF\left(
  \Li_2\left(-\frac{\tzz{r}{i}}{\tzz{i}{r}}\right)
- \ln y_{ir}\,\ln \tzz{i}{r}
\right)
\nn\\[1mm] && \qquad
- \CA \left(
  \frac12 \ln^2\frac{\tzz{i}{r}}{\tzz{r}{i}}
+ \frac12 \ln^2 y_{ir}
+ \ln y_{ir} \ln\frac{\tzz{r}{i}}{\tzz{i}{r}}
+ 2 \Li_2\left(-\frac{\tzz{r}{i}}{\tzz{i}{r}}\right)
- \frac{\pi^2}{3}\right)
\nn\\[1mm] && \qquad
+ \frac{\beta_0}{2}\ln\frac{\mu^2}{Q^2}
\Bigg)
\bra{m}{(0)}{(\momt{(ir)})}
\,\hP_{q_i g_r}^{(0)}(\tzz{i}{r},\tzz{r}{i})
\,\ket{m}{(0)}{(\momt{(ir)})}
\nn\\[1mm] && \quad
+ \CF (\CA-\CF)
\SME{m}{0}{\momt{(ir)}}
\Bigg]
\,.
\label{eq:finCqg10+finCqgRx00}
\eeeq
The case of $f_i=g$, $f_r=q$ is obtained from the latter with the
$i\leftrightarrow r$ substitution.

The finite parts of the soft subtractions are as follows:
\beeq
\finite
\aand\!\!
\Big[\cS{ir}{(0,1)}(\mom{}) + \cS{ir}{(0,0\otimes I)}(\mom{})\Big] =
- 4\as^2 \sum_i\sum_{k\ne i} \frac12 \calS_{ik}(r)
\nn\\&& \times
\Bigg[
\,2\Real \bra{m}{(0)}{(\momt{(r)})}
\,\bT_i \bT_k
\,\finite\ket{m}{(1)}{(\momt{(r)})}
\nn\\&&\quad
+ \sum_j
\Bigg(
\sum_{l\ne j}
\frac12 \finite\Big[
\IcS{jl}{(0)}(y_{\ti{j}\ti{l}}, y_{\ti{j}Q}, y_{\ti{l}Q}; m) + \IcCS{(0)}(m)
\Big]
\SME{m;(i,k)(j,l)}{0}{\momt{(r)}}
\nn\\&& \qquad\qquad\quad
+ \finite\Big[\IcC{j}{(0)}(y_{\ti{j}Q}; m) + \IcCS{(0)}(m)
\Big]
\,\bT_j^2\,\SME{m;(i,k)}{0}{\momt{(r)}}
\Bigg)
\Bigg]
\,,
\label{eq:finSir01+finSir00xI}
\eeeq
\beeq
\finite
\aand\!\!
\Big[\cS{ir}{(1,0)}(\mom{}) + \cS{ir}{R\times (0,0)}(\mom{})\Big] =
4\as^2 
\sum_i\sum_{k\ne i}
\Bigg\{
\CA 
\,\frac12 \calS_{ik}(r)
\,\SME{m;(i,k)}{0}{\momt{(r)}}
\nn\\&& \times
\Bigg(
  \frac12 \ln^2 \frac{y_{ik}}{y_{ir} y_{kr}}
- \frac{\pi^2}{3}
- \frac{\beta_0}{2\CA} \ln \frac{\mu^2}{Q^2}
- \finite
\Big[\frac12\IcC{gg}{(0)}(y_{rQ}; m) + \Nf\IcC{q\qb}{(0)}(y_{rQ}; m)\Big]
\nn\\&& \quad
- \finite
\Big[
  \IcS{ik}{(0)}(y_{ik}, y_{iQ}, y_{kQ}; m)
- \IcS{ir}{(0)}(y_{ir}, y_{iQ}, y_{rQ}; m)
- \IcS{kr}{(0)}(y_{kr}, y_{kQ}, y_{rQ}; m) 
- \IcCS{(0)}(m) \Big]
\Bigg)
\nn\\&& 
- 2\pi\sum_{l\ne i,k} \ln\frac{y_{kl}}{y_{kr} y_{lr}}
\SME{m;(i,k,l)}{0}{\momt{(r)}}
\Bigg\}
\,.
\label{eq:finSir10+finSirRx00}
\eeeq

Finally, we present the finite parts of the collinear-soft subtractions:
\beeq
\finite
\aand\!\!
\Big[\cCS{ir}{r}{(0,1)}(\mom{}) + \cCS{ir}{r}{(0,0\otimes I)}(\mom{})\Big] =
4\as^2 \frac{2}{s_{ir}}\frac{\tzz{i}{r}}{\tzz{r}{i}}\,\bT_i^2
\nn\\&& \times
\Bigg[
\,2\Real \bra{m}{(0)}{(\momt{(r)})}
\,\finite\ket{m}{(1)}{(\momt{(r)})}
\nn\\&& \qquad
+ \sum_j
\Bigg(
\sum_{l\ne j}
\finite\Big[
\IcS{jl}{(0)}(y_{\ti{j}\ti{l}}, y_{\ti{j}Q}, y_{\ti{l}Q}; m) + \IcCS{(0)}(m)
\Big]
\SME{m;(j,l)}{0}{\momt{(r)}}
\nn\\&& \qquad\qquad\quad
+ \finite\Big[\IcC{j}{(0)}(y_{\ti{j}Q}; m) + \IcCS{(0)}(m)
\Big]
\,\bT_j^2\,\SME{m}{0}{\momt{(r)}}
\Bigg)
\,,
\label{eq:finCirSr01+finCirSr00xI}
\eeeq
\beeq
\finite
\aand\!\!
\Big[\cCS{ir}{r}{(1,0)}(\mom{}) + \cCS{ir}{r}{R\times (0,0)}(\mom{})\Big] =
- 4\as^2\,\CA\,\frac{2}{s_{ir}}\frac{\tzz{i}{r}}{\tzz{r}{i}}
\,\bT_i^2 \SME{m}{0}{\momt{(r)}}
\nn\\&& \times
\Bigg(
  \frac12 \ln^2  \frac{\tzz{i}{r}}{y_{ir} \tzz{r}{i}}
- \frac{\pi^2}{3}
- \frac{\beta_0}{2\CA} \ln \frac{\mu^2}{Q^2}
- \finite
\Big[\frac12\IcC{gg}{(0)}(y_{rQ}; m) + \Nf\IcC{q\qb}{(0)}(y_{rQ}; m)\Big]
\nn\\&& \qquad
+ \finite
\Big[ \IcS{ir}{(0)}(y_{ir}, y_{iQ}, y_{rQ}; m) + \IcCS{(0)}(m) \Big]
\Bigg)
\,.
\label{eq:finCirSr10+finCirSrRx00}
\eeeq

For the sake of completeness, we recall the finite part of the first two
terms in \eqn{eq:sigmaNNLOm+1} from \Ref{Somogyi:2006_1} adapted to the
present case: 
\beeq
&&
\left[\dsig{RV}_{m+1} + \dsig{R}_{m+1} \otimes \bI(m,\eps)\right]_{\eps = 0} =
{\cal N}\,\frac{\as}{2\pi} \sum_{\{m+1\}}
\PS{m+1}{}\frac{1}{S_{\{m+1\}}}
\nn\\&&\times
\Bigg\{
2 \Real \bra{m+1}{(0)}{(\mom{})}\finite \ket{m+1}{(1)}{(\mom{})}
\nn\\&&\quad\qquad
+ \sum_{i} \Bigg[ \sum_{k\ne i}
  \finite\left[\IcS{ik}{}(y_{ik},y_{iQ},y_{kQ};m) + \IcCS{}(m)\right]
\,\SME{m+1;(i,k)}{0}{\mom{}}
\nn\\&&\qquad\qquad\qquad
+ \finite\left[\IcC{i}{}(y_{iQ};m)  + \IcCS{}(m)\right]\,\bT_i^2
\,\SME{m+1}{0}{\mom{}}
\Bigg]
\Bigg\}
\,.
\qquad~
\eeeq

\end{document}